%% file: 00-main.tex
\RequirePackage{fix-cm}
\documentclass[twocolumn,epjc3]{svjour3}
\smartqed  
\RequirePackage{graphicx}
\usepackage{epsfig}
\usepackage{bm} 
\usepackage[usenames]{color}
\usepackage{bbding}
\usepackage{slashed}
\newcommand{\blue}[1]{\textcolor{black}{#1}}

\newcommand{\Tr}{\mbox{Tr}}
\newcommand{\vphi}{\varphi}
\newcommand{\Lagr}{{\cal L}}
\newcommand{\cO}{{\cal O}}
\newcommand{\Omslash}{\!\not{\!\! \omega}}
\newcommand{\OmBarslash}{\!\not{\!\! \bar{\omega}}}

\newcommand{\bgs}{B\rightarrow X_s\gamma}

\newcommand{\gap}{\mbox{\hspace{0.5cm}}}
\newcommand{\pard}{\partial}
\newcommand{\BB}{\mbox{\boldmath$B$}}
\newcommand{\BW}{\mbox{\boldmath$W$}}
\newcommand{\BV}{\mbox{\boldmath$V$}}
\newcommand{\Bpard}{\mbox{\boldmath$\partial$}}
\newcommand{\dslash}{\!\not{\! \partial}}
\newcommand{\Bdslash}{\!\not{\! \Bpard}}
\newcommand{\Aslash}{\!\not{\!\! A}}
\newcommand{\Wslash}{\!\not{\!\! W}}
\newcommand{\BWslash}{\!\not{\!\! \BW}}
\newcommand{\Bslash}{\!\not{\!\! B}}
\newcommand{\BBslash}{\!\not{\!\! \BB}}
\newcommand{\Zslash}{\!\not{\!\! Z}}
\newcommand{\BVslash}{\!\not{\! \BV}}
\newcommand{\cL}{{\cal L}}
\newcommand{\cM}{{\cal M}}
\newcommand{\veps}{\varepsilon}

\newcommand{\diag}{\mbox{diag}}

\newcommand{\unit}[1]{\;\mathrm{#1}}
\journalname{Eur. Phys. J. C}
\begin{document}

\title{The vector resonance triplet with the direct coupling to the third quark generation
}


\author{Mikul\'{a}\v{s} Gintner\thanksref{e1,addr1,addr2}
        \and
        Josef Jur\'{a}\v{n}\thanksref{e2,addr2,addr3} 
}

\thankstext{e1}{e-mail: gintner@fyzika.uniza.sk}
\thankstext{e2}{e-mail: josef.juran@utef.cvut.cz}

\institute{Physics Department, University of \v{Z}ilina,
Univerzitn\'{a} 1, 010 26 \v{Z}ilina, Slovakia \label{addr1}
           \and
           Institute of Experimental and Applied Physics,
Czech Technical University in Prague, Horsk\'{a} 3a/22, 128 00
Prague, Czech Republic \label{addr2}
           \and
           Institute of Physics, Silesian University in Opava,
Bezru\v{c}ovo n\'{a}m. 13, 746 01 Opava, Czech Republic
\label{addr3} }

\date{Received: date / Accepted: date}

\maketitle

\begin{abstract}
The effective Lagrangian with scalar and vector resonances that
might result from new strong physics beyond the SM is formulated
and studied. In particular, the scalar resonance representing the
recently discovered 125-GeV boson is complemented with the
$SU(2)_{L+R}$ triplet of hypothetical vector resonances. Motivated
by experimental and theoretical considerations, the vector
resonance is allowed to couple directly to the third quark
generation only. The coupling is chiral-dependent and the
interaction of the right top quark can differ from that of the
right bottom quark. To estimate the applicability range of the
effective Lagrangian the unitarity of the gauge boson scattering
amplitudes is analyzed. The experimental fits and limits on the
free parameters of the vector resonance triplet are investigated.
\end{abstract}

\section{Introduction}
\label{sec:Intro}

\input{01-intro}

\section{The effective Lagrangian}
\label{sec:tBESS}

\input{02-tBESS}

\section{Tree-level unitarity limits}
\label{sec:Ulimits}

\input{03-Ulimits}

\section{The low-energy analysis}
\label{sec:LEanalysis}

\subsection{Integrating out the vector resonance triplet}
\label{subsec:IntegOut}

\input{04a-integrating_out}

\subsection{Predictions
for the low-energy observables}
\label{subsec:LEtBESSpredictions}

\input{04b-LEtBESSpredic}

\subsection{Fits and limits}
\label{subsec:FitsLimits}

\input{04c-fits_and_limits}

\section{Conclusions}
\label{sec:conclusions}

\input{05-conclusions}

\begin{acknowledgements}
We would like to thank C.~Grojean, F.~Riva, S.~Pokorski,
G.~Panico, and A.~Sopczak for useful discussions. The work was
supported by the Research Program MSM6840770029 and by the project
of International Cooperation ATLAS-CERN LG13009. J.J.\ was also
supported by the NSP grant of the Slovak Republic. M.G.\ was
supported by the Slovak CERN Fund. We would also like to thank the
Slovak Institute for Basic
Research for their support.\\
\end{acknowledgements}

\appendix

\input{appendix-A}
\input{appendix-B}

\input{biblio}

\end{document}

%% file: 01-intro.tex
The ATLAS and CMS announcements of the 125-GeV boson
discovery~\cite{125GeVBosonDiscovery}
have provided major
contribution towards finding the solution of the puzzle about the
mechanism of electroweak symmetry breaking (ESB). The recent data
revelations and analyses~\cite{Moriond2013} strongly suggest that
the observed 125-GeV boson is a Higgs-like particle with a tight
relationship with the ESB mechanism.  Nevertheless, the question
about the true nature of the mechanism, and thus about physics
beyond the Standard model (SM), remains unsolved. While the
observed properties of the discovered boson are compatible with
the SM Higgs boson
hypothesis~\cite{EllisYou,HiggsAtLast,NBlikeSMhiggs,GlobalSMfit},
at the same time they are compatible with some alternative
extensions of the SM~\cite{CapabilityOfMeasurements}.

From a theoretical point of view, the alternatives to the SM Higgs
get some preferences due to the naturalness argument.
The extensions of the SM still in the game include
theories where
electroweak symmetry is broken by new strong interactions, like in
Technicolor~\cite{TC,ETC,WalkingTC,TopcolorTC}.

Most studies aimed at the evaluation of the impact of the new
discovery on the alternatives theories assume the boson has a spin
zero. This assumption gets a growing experimental support as more
LHC data is being processed. Of course, at the same time it
disfavors strongly-interacting theories without a light scalar
field and calls for theories with a light composite
strongly-interacting
Higgs~\cite{CompositeScalar-old,CHsketch,CompositeScalar,HiggsLag,CompositeScalar-GGduality,alternatives}
of a proper mass.

A feverish activity in building effective descriptions and
identifying possible underlying theories is taking place on the
theoretical front nowadays. The focus lies in the modeling,
parameterizing, and fitting the 125-GeV Higgs-like boson sector of
candidate theories (see, e.g.,
\cite{ContinoEffLagrForLightHiggs}). Effectively, the Higgs-like
boson can be described as a stand-alone singlet added to the
non-linear sigma model of the Nambu--Goldstone bosons. This is the
most model-independent approach, but with the least predictive
power. Alternatively, the Higgs-like scalar \blue{can be} made a member of a
multiplet of the symmetry of the strong
sector~\cite{HiggsInMultiplet}. The latter approach results in
additional experimentally testable restrictions on free parameters
of the model. At the same time, it can provide a mechanism for
keeping the scalar resonance light.

Following theoretical arguments, as well as the example of QCD, it
seems reasonable to expect that beside the composite scalar the
new strong interactions would also produce bound states of higher
spins. A natural candidate to look for is the vector $SU(2)$
triplet resonance. From another point of view, if
the composite Higgs couplings differ from the SM ones,
as is usually the case in strongly interacting theories,
the Higgs alone will fail to unitarize the $VV$ $(V=W^\pm,Z)$
scattering amplitudes.
Then, additional resonances are required to tame the unitarity.

In this paper, we study the effective Lagrangian where beside the 125-GeV scalar
resonance --- an $SU(2)_{L+R}$ singlet complementing the non-linear triplet
of the Nambu--Goldstone bosons --- the $SU(2)_{L+R}$ triplet of vector
resonances is explicitly present. It fits the situation when the global
$SU(2)_L\times SU(2)_R$ symmetry is broken down to $SU(2)_{L+R}$.
As far as the vector resonance sector is concerned,
the vector triplet is introduced as a
gauge field via the hidden local symmetry approach~\cite{HLS}.
Because of this, the vector resonance mixes with the EW gauge bosons,
which results in appearance of indirect couplings of the vector resonance
with all SM fermions. Besides, the direct couplings of the vector resonance
triplet to the SM fermions are also allowed by the Lagrangian symmetry.
Regarding the direct couplings we opt for a special setup
inspired by the speculations about an extraordinary role of the top quark
(or the third quark generation) in new strong physics: we
admit direct couplings of the new triplet to no other SM fermions, but
the top and bottom quarks only.
Finally, the symmetry allowed interaction terms between the scalar
and vector resonances are also present.

In the strong scenario, the direct couplings between the SM
fermions and the vector resonance can depend on the degree of
compositeness of a given fermion as well as on symmetry group
representations the fermions are organized into. In principle, the
degree of compositeness of the SM fermions can vary for different
flavors and chirality. Theories that can be related to our
effective description include 2-site deconstructed models, purely
4-dimensional multi-site models, and composite Higgs
models~\cite{CompositeScalar-GGduality,HiggsInMultiplet,ContinoResonancesInCmpHiggs,Sundrum,DeCurtis,PanicoPrivate}.
All these models predict the existence of resonances of higher
spins, including the vectorial ones. The idea of partial
compositeness that appears in some of these models could justify
the exclusivity of the third quark direct couplings to the vector
resonances in our effective Lagrangian.

The couplings of the hypothetical vector resonance to light
fermions are tightly restricted by the existing measurements from
the LEP, SLC, and Tevatron experiments. Thus, it is reasonable to
neglect them also from the experimental point of view.
The direct coupling of the vector resonance to the bottom quark is
also restricted by the experiments through the measurement of the
$Zbb$ vertex, at least. In our effective Lagrangian, the influence
of this restriction on the direct interaction with the top quark
has been weakened by the splitting of the interaction with the
right top and bottom quark.

As far as the direct LHC bottom limits on the vector resonance masses
are concerned they are strongly model-dependent. In general,
it can be said that considering the partial compositeness
for the third quark generation only admits the limits to be as low
as 300 GeV, or even less, for certain values of the Higgs-like
boson couplings~\cite{CHsketch}. The most restrictive bottom limit
is obtained when no compositeness of the SM fermions is assumed;
it is slightly below $1\unit{TeV}$.

In this paper, we study the unitarity constraints and the best
fits of the vector resonance free parameters to the existing data.
We perform the best-fit analysis under the simplifying assumption
that the scalar resonance couplings are the SM ones. It should
serve as an approximation of the situation allowed by the
experiment when the actual scalar couplings do not differ too much
from the SM ones. \blue{In setting constraints on the vector
resonance couplings the published LHC analyses cannot compete with the
low-energy measurements yet.} Therefore we focus on the
low-energy data when calculating the limits. Our analysis have
been performed as a multi-observable $\chi^2$-fit taking into
account the correlations among the observables used. The list of
fitted observables is comprised of $\epsilon_1$, $\epsilon_2$,
$\epsilon_3$, $\Gamma_b(Z\rightarrow b\bar{b}+X)$, and BR$(\bgs)$.
Throughout the analysis the mass of the considered vector triplet
assumes TeV values.

This paper is organized as follows. In Section~\ref{sec:tBESS}
we introduce our effective Lagrangian.
In Section~\ref{sec:Ulimits} the tree-level unitarity limits
for the longitudinal electroweak gauge boson scattering
as function of the scalar and vector resonance parameters are
calculated and discussed.
Section~\ref{sec:LEanalysis}
is devoted to the low-energy analysis of the vector resonance couplings.
Section~\ref{sec:conclusions} contains our conclusions followed by appendices.

%% file: 02-tBESS.tex

We introduce the $SU(2)_{L+R}$ triplet vector resonance to
the usual $SU(2)_L\times SU(2)_R\rightarrow SU(2)_{L+R}$
effective Lagrangian with the non-linearly transforming
$SU(2)_{L+R}$ triplet of the would-be Nambu--Goldstone bosons
augmented with the $SU(2)_{L+R}$ singlet scalar resonance.
The vector triplet is brought in as a gauge field via the
hidden local symmetry (HLS) approach~\cite{HLS}. The effective Lagrangian
is built to respect the global $SU(2)_L\times SU(2)_R\times U(1)_{B-L}\times
SU(2)_{HLS}$ symmetry of which the $SU(2)_L\times
U(1)_Y\times SU(2)_{HLS}$ subgroup is also a local symmetry.
The $SU(2)_{HLS}$ symmetry is an auxiliary gauge symmetry
invoked to accommodate the $SU(2)$
triplet of vector resonances. Beside the scalar singlet and
the vector triplet, the effective Lagrangian is built out of
the SM fields only.

Our effective Lagrangian can be split in three terms
\begin{equation}
  \cL = \cL_\mathrm{GB} + \cL_\mathrm{ESB} + \cL_\mathrm{ferm},
  \label{eq:tBESSLag}
\end{equation}
where $\cL_\mathrm{GB}$ describes the gauge boson sector including
the $SU(2)_\mathrm{HLS}$ triplet, $\cL_\mathrm{ESB}$ is the scalar sector
responsible for spontaneous breaking of the electroweak and
hidden local symmetries, and $\cL_\mathrm{ferm}$ is the fermion Lagrangian
of the model.

Beside the SM gauge fields $W_\mu^a(x)$ and $B_\mu(x)$,
the $SU(2)_\mathrm{HLS}$ gauge triplet
$\vec{V}_\mu=(V_\mu^1,V_\mu^2,V_\mu^3)$ represents hypothetical
neutral and charged vector resonances of a new strong sector.
Under the
$[SU(2)_L\times SU(2)_R]^\mathrm{glob}\times SU(2)_\mathrm{HLS}^\mathrm{loc}$
group the triplet transforms as
\begin{equation}
  \BV_\mu \rightarrow h^\dagger \BV_\mu h +h^\dagger\pard_\mu h,
\end{equation}
where $h(x)\in SU(2)_\mathrm{HLS}^\mathrm{loc}$ and $\BV_\mu =
i\frac{g''}{2}V_\mu^a\tau^a$. The $2\times 2$ matrices
$\vec{\tau}=(\tau^1,\tau^2,\tau^3)$ are the $SU(2)$ generators.

The ESB sector
contains six unphysical real scalar fields, would-be Goldstone
bosons of the model's spontaneous symmetry breaking.
The six real scalar fields
$\vphi_L^a(x), \vphi_R^a(x),\; a=1,2,3$, are introduced as
parameters of the~$SU(2)_L\times SU(2)_R$ group elements in
the exp-form
$\xi(\vec{\vphi}_{L,R})=\exp(i\vec{\vphi}_{L,R}\vec{\tau}/v)\in SU(2)_{L,R}$
where $\vec{\vphi}=(\vphi^1,\vphi^2,\vphi^3)$.

In the ESB sector the would-be Goldstone bosons couple
to the gauge bosons $W_\mu^a$, $B_\mu$, $V_\mu^a$,
and to the 125-GeV scalar resonance $h(x)$ obeying
$[SU(2)_L\times U(1)_Y\times SU(2)_\mathrm{HLS}]^\mathrm{loc}$
symmetry requirements. Thus,
\begin{eqnarray}\label{eq:LagrESB}
  \Lagr_\mathrm{ESB} &=&  \frac{1}{2}\pard_\mu h\,\pard^\mu h
                         -\frac{1}{2}M_h^2 h^2
  \nonumber\\
                &&-v^2\left[\Tr\left(\bar{\omega}_\mu^\perp\right)^2
                +\alpha\;\Tr\left(\bar{\omega}_\mu^\parallel\right)^2\right]
  \nonumber\\
  && \phantom{-}\;\times\,(1+2a\frac{h}{v}+a'\frac{h^2}{v^2}+\ldots),
\end{eqnarray}
where $\alpha$, $a, a', \ldots$ are free parameters, $M_h= 125\unit{GeV}$,
and $\bar{\omega}_\mu^{\parallel,\perp}$
are, respectively, $SU(2)_{L-R}$ and $SU(2)_{L+R}$ projections of the gauged
Maurer--Cartan 1-form,
\begin{eqnarray}
   \bar{\omega}_\mu^{\parallel} &=& \omega_\mu^{\parallel}+
   \frac{1}{2}\left(\xi_L^\dagger\BW_\mu\xi_L+\xi_R^\dagger\BB_\mu\xi_R\right)-
   \BV_\mu,
   \label{eq:gaugeMCparallel}\\
   \bar{\omega}_\mu^{\perp} &=& \omega_\mu^{\perp}+
   \frac{1}{2}\left(\xi_L^\dagger\BW_\mu\xi_L-\xi_R^\dagger\BB_\mu\xi_R\right),
   \label{eq:gaugeMCperp}
\end{eqnarray}
where $\omega_\mu^{\parallel,\perp}=
(\xi_L^\dagger\pard_\mu\xi_L\pm\xi_R^\dagger\pard_\mu\xi_R)/2$.
When $a=a'=1$ and all other $a$'s
are zeros, the scalar resonance imitates the SM Higgs boson.

The masses of the vector triplet are set by the scale $v$ and
depend on the three gauge couplings $g, g', g''$, and the free
parameter $\alpha$. In the limit when $g$ and $g'$ are negligible
compared to $g''$, the masses of the neutral and charged resonances
are degenerate, $M_{V} = \sqrt{\alpha}g''v/2$. If higher order
corrections in $g/g''$ are admitted, a tiny mass splitting occurs
such that $M_{V^0}>M_{V^\pm}$~\cite{tBESS}.

As far as the fermion sector is concerned no new fermions
beyond the SM have been introduced in our Lagrangian.
The fermion sector of the Lagrangian can be divided into three parts
\begin{equation}\label{eq:tBESSLagrFerm}
  \cL_\mathrm{ferm} = \cL_\mathrm{ferm}^\mathrm{SM}
                      + \cL_\mathrm{ferm}^\mathrm{scalar}
                      + \cL_{(t,b)}^\mathrm{tBESS},
\end{equation}
where $\cL_\mathrm{ferm}^\mathrm{SM}$ contains
the SM interactions of fermions with the electroweak gauge bosons,
$\cL_\mathrm{ferm}^\mathrm{scalar}$ is about the interactions
of the fermions with scalar fields and includes the fermion masses,
and $\cL_{(t,b)}^\mathrm{tBESS}$ describes the third quark generation
direct interactions with the vector resonance. In addition, it
contains symmetry allowed non-SM interactions of
the third quark generation with the EW gauge bosons.

The first term of~(\ref{eq:tBESSLagrFerm}) is identical
to its SM counterpart. Namely,
\begin{equation}
  \cL_\mathrm{ferm}^\mathrm{SM} = \sum_{\forall\psi}
  \left[ I_c^L(\psi_L)+I_c^R(\psi_R) \right],
\end{equation}
where $\psi$ denotes the usual $SU(2)$ doublets\footnote{
   Of course, the $SU(2)_R$ symmetry
   is broken by the weak hypercharge interactions and, thus,
   the $SU(2)_R$ fermion doublets are not well justified
   once the global symmetry gets gauged.
} of SM fermions and the sum runs through them.
The invariants $I_c^{L,R}$ read
\begin{eqnarray}
  I_c^L(\psi_L) &=& i\bar{\psi}_L(\Bdslash+\BWslash+\BBslash) \psi_L,
  \\
  I_c^R(\psi_R) &=& i \bar{\psi}_R(\Bdslash+\BBslash) \psi_R.
\end{eqnarray}

The second term of the Lagrangian~(\ref{eq:tBESSLagrFerm}) reads
\begin{equation}\label{eq:LagrFermScalar}
  \cL_\mathrm{ferm}^\mathrm{scalar} =
  -\sum_i I_\mathrm{mass}(\psi^i)\,
  (1+c_i\frac{h}{v}+c_i'\frac{h^2}{v^2}+\ldots),
\end{equation}
where $c_i, c_i', \ldots$ are free parameters, and
\begin{equation}\label{eq:FermionMassTerm}
  I_\mathrm{mass}(\psi^i) = \bar{\psi}_L^i U M_f^i \psi_R^i + \mbox{H.c.},
\end{equation}
where $M_f^i$ is a $2\times 2$ diagonal matrix with the masses of
the upper and bottom $i$th fermion doublet components on its diagonal,
and $U=\xi(\vec{\pi})\cdot\xi(\vec{\pi})=\exp(2i\vec{\pi}\vec{\tau}/v)$.
Note that when $c_i=1, \forall i$, and the rest of $c$'s
are zeros the scalar resonance interactions with fermions imitate
the corresponding interactions of the SM Higgs boson.

The third term of~(\ref{eq:tBESSLagrFerm}) coincides with
the corresponding part of the Lagrangian that we introduced in~\cite{tBESS}.
The effective Lagrangian in~\cite{tBESS} was a Higgs-less description
of a vector resonance triplet that was made obsolete by the 125-GeV boson
discovery. Nevertheless, the motivation for the vector resonance interaction
pattern with fermions that was used in~\cite{tBESS} has remained unchanged
and we use the same pattern in this paper. Thus, the vector resonance
couples directly to the third quark generation only. The interactions of
the left and right fields are proportional to $b_L$ and $b_R$, respectively.
In addition, there is a free parameter $p$ which disentangles the right
bottom coupling from the right top coupling. The assumption that
the vector resonance interaction with the right bottom quark is weaker than the
interaction with the right top quark corresponds to the expectation that
$0\leq p \leq 1$. While $p=1$ leaves the interactions equal, the $p=0$ turns off the
right bottom quark interaction completely and maximally breaks the
$SU(2)_R$ part of the Lagrangian symmetry down to $U(1)_{R3}$.
In addition, the symmetry of the Lagrangian admits non-SM interaction
of the fermions with the EW gauge bosons that we also include
in $\cL_{(t,b)}^{\mbox{\scriptsize tBESS}}$ under the assumption that
they apply to the third quark generation only. These interactions
are proportional to the free parameters $\lambda_L$ and $\lambda_R$.
The $\cL_{(t,b)}^\mathrm{tBESS}$ Lagrangian reads\footnote{
Throughout this paper we use the `tBESS' label
to indicate the $SU(2)_{L+R}$ vector resonance triplet
with this particular interaction pattern to fermions.
The label is inspired by the fact that our vector resonance
in our effective Lagrangian is introduced in the same way as in
the BESS model~\cite{BESS} and that \textbf{t}op quark and/or
\textbf{t}hird quark
generation has a special standing in its interactions,
different from the original BESS model.}
\begin{eqnarray}
  \cL_{(t,b)}^{\mbox{\scriptsize tBESS}} &=&
  b_L\left[ I_b^L(\psi_L)-I_c^L(\psi_L) \right]
  \nonumber\\
  &&  +b_R\left[ I_b^R(P\psi_R)-I_c^R(P\psi_R) \right]
  \nonumber\\
  && +2\lambda_L I_\lambda^L(\psi_L) +2\lambda_R I_\lambda^R(P\psi_R),
  \label{eq:LagrFermTBESS}
\end{eqnarray}
where $\psi=(t,b)$.
The invariants $I_b^{L,R}$ and $I_\lambda^{L,R}$ read
\begin{eqnarray}
  I_b^h(\psi_h) &=& i\bar{\chi}_h\left[\Bdslash+\BVslash+ig'\Bslash (B-L)/2\right]\chi_h,
  \\[0.2cm]
  I_\lambda^h(\psi_h) &=& i\bar{\chi}_h \OmBarslash^\perp \chi_h
  \nonumber\\
  &=& i \bar{\chi}_h\left[\;\Omslash^\perp+
      (\xi_L^\dagger\BWslash \xi_L-\xi_R^\dagger\BBslash^{R3} \xi_R)/2\right]\chi_h,\;\;\;\;\;\;
\end{eqnarray}
where $B$ and $L$ are the baryon and lepton number operators, respectively, $h=L,R$,
$\chi_h \equiv \chi(\vec{\vphi}_h,\psi_h) = \xi^\dagger(\vec{\vphi}_h)\cdot\psi_h$,
$\BBslash^{R3}=ig'\slashed{B}\tau^3$.
The matrix $P=\diag(1,p)$
disentangles the direct interaction of the vector triplet with the
right top quark from the interaction with the right bottom quark.

Note that under the parity transformation, $I_b^L\leftrightarrow I_b^R$
and $I_\lambda^L\leftrightarrow -I_\lambda^R$. Therefore, the new physics
interactions in the fermion Lagrangian break parity, unless $p=1$,
$b_L=b_R$, and $\lambda_L=-\lambda_R$.

The direct couplings of the top and bottom quarks can be due to
their partial compositeness. They can emerge from the underlying theory
through interweaving the top and bottom quark fields with
the fermionic operators in the new strong sector.
The absence of the vector resonance direct couplings
with light SM fermions can indicate that
the fermions are elementary.

In the unitary (physical) gauge where all six unphysical scalar
fields are gauged away the gauged MC 1-form projections (\ref{eq:gaugeMCparallel})
and (\ref{eq:gaugeMCperp}) read
\begin{eqnarray}
  \bar{\omega}^\perp_{\mu} = \frac{1}{2}(\BW_\mu-\BB_\mu), \\
  \bar{\omega}^\parallel_{\mu} = \frac{1}{2}(\BW_\mu+\BB_\mu)-\BV_\mu.
\end{eqnarray}
Thus, the ESB Lagrangian~(\ref{eq:LagrESB}) assumes the form
\begin{eqnarray}\label{eq:LagrESBgaugeU}
  \Lagr_\mathrm{ESB} &=&  \frac{1}{2}\pard_\mu h\,\pard^\mu h
                         -\frac{1}{2}M_h^2 h^2
         -\frac{v^2}{4}\left\{\Tr\left(\BW_\mu-\BB_\mu\right)^2\right.
  \nonumber\\
   &&+\left.\alpha\;\Tr\left[\left(\BW_\mu+\BB_\mu\right)-2\BV_\mu\right]^2\right\}
  \nonumber\\
  && \phantom{-}\;\times\,\left(1+2a\frac{h}{v}+a'\frac{h^2}{v^2}+\ldots\right).
\end{eqnarray}
In the fermion sector the Lagrangian~(\ref{eq:LagrFermScalar}) turns into
\begin{equation}\label{eq:LagrFermScalarGaugeU}
  \cL_\mathrm{ferm}^\mathrm{scalar} =
  -\frac{1}{v}\sum_i \left(\bar{\psi}_L^i M_f^i\psi_R^i\right)
  \left(1+c_i\frac{h}{v}+c_i'\frac{h^2}{v^2}+\ldots\right)
\end{equation}
and the new physics part of the $(t,b)$ Lagrangian
assumes the form
\begin{eqnarray}
  \cL_{(t,b)}^\mathrm{tBESS} &=&
                            i b_L \bar{\psi}_L(\BVslash-\BWslash) \psi_L
  \nonumber\\
                 && + i b_R \bar{\psi}_R P(\BVslash-\BBslash^{R3}) P\psi_R
  \nonumber\\
  & &  + i \lambda_L \bar{\psi}_L(\BWslash-\BBslash^{R3}) \psi_L
  \nonumber\\
  & &  + i \lambda_R \bar{\psi}_RP(\BWslash-\BBslash^{R3}) P\psi_R.
  \label{eq:LagrFermTBESSinUgauge}
\end{eqnarray}

To obtain the masses of the electroweak gauge bosons as well as of
the new vector resonances their mass matrix has to be
diagonalized.
The unitary gauge ESB Lagrangian~(\ref{eq:LagrESBgaugeU}) expressed in
the gauge boson mass basis reads
\begin{eqnarray}\label{eq:LagrESBgaugeUMassBasis}
  \Lagr_\mathrm{ESB} &=&  \phantom{+}\frac{1}{2}\pard_\mu h\,\pard^\mu h
                         -\frac{1}{2}M_h^2 h^2
 \nonumber\\
 && +\frac{1}{2}\left(M_Z^2Z_\mu Z^\mu+2M_W^2W^+_\mu W^{-\mu} \right.
 \nonumber\\
 && \left.
    \quad\quad\ + M_{V^0}^2V^0_\mu V^{0\mu}+2M_{V^{\pm}}^2V^+_\mu V^{-\mu}\right)
 \nonumber\\
  && \phantom{-}\;\times\,\left(1+2a\frac{h}{v}+a'\frac{h^2}{v^2}+\ldots\right).
\end{eqnarray}

Once the gauge boson fields are expressed in the gauge boson mass
basis, the mixing generated interactions of the vector
triplet with all fermions will emerge from the fermion Lagrangian
$\cL_\mathrm{ferm}^\mathrm{SM}$.
However, these indirect interactions of the vector resonance
with the light fermions will be suppressed
by the mixing matrix elements proportional to $1/g''$.

The request that our Lagrangian be treatable perturbatively bounds
the values of $g''$ from above by the naive perturbativity limit,
$g''/2\stackrel{<}{\sim} 4\pi$, \blue{implying $g''\stackrel{<}{\sim}25$.
If we took this value as the final say in this issue it would not be
reasonable to use $g''$ higher than about 20 in our calculations.
Nevertheless, one can imagine that a more rigorous analysis of 
the perturbativity limit could somehow modify its value one way or the other. 
For this reason, as well as motivated by the best fit value
of $g''= 29$ (see the Eq.~(\ref{eq:BestFit1}) in Subsection~\ref{subsec:FitsLimits}), 
we show results of our analysis for $g''$ up to 30.
Nevertheless, the reader should keep in mind that the chance that
the shown results are not meaningful grows with $g''$, especially above 20.}

\blue{
In principle, the vector resonance parameters can be constrained even before its discovery
through measurements of observables affected by the resonance existence.
For example, the measurement of the gauge boson self-interactions
can be used to restrict the coupling $g''$ of our effective Lagrangian.
In particular, the triple gauge boson couplings (TGC) were probed by various
experiments: D0, LEP, ATLAS, and CMS~\cite{TGCD0,TGCLEP,TGCATLAS,TGCCMS}. 
Among these, the most stringent constraints originate from the LEP measurement of 
$W$-pairs in $e^+e^-\rightarrow W^+W^-$.
Using the results of the analysis of the LEP data in~\cite{TGCLEP,TGCATLAS} we get 
the lower bound on $g''\geq 5$ at $95\%$~CL ($\Delta\kappa_Z\geq -0.02$).
In~\cite{TGCcombined} the TGC coupling constraints
were obtained by combining the Higgs data with 
the D0, LEP, and ATLAS measurements.
Their results ($\Delta\kappa_Z\geq -0.004$) imply
the lower bound on $g''\geq 11$ at $95\%$~CL. 
Note that the combined lower limit on $g''$ converges on 
the lower $95\%$~CL limit, $g''\geq 12$, that will be obtained in 
Subsection~\ref{subsec:FitsLimits} from the low-energy data.
However, the reader should be warned that while the limits for
$\Delta\kappa_Z$ were derived in the formalism with three free
parameters the tBESS TGC coupling depends on a single free parameter.
The rigorous derivation of the $g''$ limit
would require some additional constraints
to be imposed in the TGC analysis from the onset.
Thus, the TGC implied $g''$ limits shown above should be taken as
estimates only.
}

\blue{
The ATLAS constraints on the $Wtb$ vertex~\cite{WtbATLAS} can be used
to derive limits on the vector resonance couplings to fermions.
The limits are shown in footnote~\ref{fn1}, 
Subsection~\ref{subsec:FitsLimits}, where they can be confronted
with the low-energy constraints obtained therein. Let us advertise that
the $Wtb$ induced limits are not competitive yet.
Unfortunately, the assumptions used in the newer analysis 
of the $Wtb$ vertex by the CMS~\cite{WtbCMS}, as well as in the combined
ATLAS+CMS analysis~\cite{WtbCombined}, are not compatible with
our formalism.
}

\blue{
There are other LHC measurements that are candidates for restricting 
the vector resonance parameters. At this point, however,
they do not provide useful restriction on the tBESS parameters because
either an appropriate analysis of needed observables is missing 
or there is no sufficient statistics yet.
}

As far as the scalar parameters are concerned
the direct LHC measurements restrict $a$ and
$c_i$'s to the vicinity of their SM
values~\cite{EllisYou,HiggsAtLast,LHCdataAndre,newHiggsData}.
In particular,
authors of~\cite{HiggsAtLast} calculated constraints on seven free parameters
of the effective Lagrangian with the 126 GeV scalar.
Using the most recent LHC Higgs data
in all available search channels in combination with electroweak precision
observables from SLC, LEP-1, LEP-2, and the Tevatron.
they found a restriction $0.98\leq a\leq 1.08$ at $95\%$CL
when the seven-parameter fit was performed.
Under the assumptions inspired by the composite Higgs scenario when
the Higgs couples equally to all fermions and the NLO couplings are set to zero
the parameter $a$ is restricted to be within 10\% of the SM value at $95\%$~CL.
Regarding the $c_i$ parameters $c_t$, $c_b$, and $c_\tau$ only
have been restricted by the LHC measurements so far.
The composite Higgs scenario fit implies $0.7\leq c_t=c_b=c_\tau\leq
1.2$ at $95\%$~CL~\cite{HiggsAtLast}.

In this paper the analysis of our effective Lagrangian
focuses on the setting unitarity restrictions for the validity of
the Lagrangian and on the fitting of the Lagrangian free parameters
using the low-energy precision data. As it was argued in~\cite{HiggsAtLast}
the contributions of the Lagrangian terms proportional to $h^2$,
or higher powers of $h$, can be neglected in the electroweak precision
analysis. At the same time, these terms have not been probed by the existing data.
Thus, in our analysis we can ignore all these terms.
As far as $c_i$'s are concerned the electroweak observables are not
sensitive to them at one loop level, neither they influence our unitarity
calculations.

Having said that the fitting calculations will be performed under
simplifying assumptions $a=c_i=1$, i.e.\ the scalar resonance couples
as the SM Higgs boson. This assumption can be regarded as an approximation of
the situation when the parameters do not differ too much from their SM values.
The assumption is made for the sake of simplification of the analysis
when our major goal is to get a basic picture about the interplay between
the scalar and vector resonances in the effective Lagrangian.
A more general study for non-SM scalar resonance couplings is in progress.

%% file: 03-Ulimits.tex

The SM without the SM Higgs boson is not renormalizable and its
amplitudes violate unitarity at some energy. In particular,
if the couplings of the 125-GeV Higgs differ from those of
the SM Higgs boson they fail in unitarization of the gauge
boson scattering amplitudes. In this situation, the introduction
of other resonances might be necessary in order to fix
the unitarity or to postpone its violation, at least.

In the Higgs-less SM, one can estimate the scattering energy at
which the unitarity violation occurs using the Equivalence Theorem
\cite{EquivalenceTheoremRenorm,EquivalenceTheoremNonRenorm}
approximation of the $W_L^+W_L^-$, $Z_L Z_L$, $W_L^\pm Z_L$, and
$W_L^\pm W_L^\pm$ scattering by the pionic scattering amplitudes
of the non-linear sigma model. Thus, one can find that the SM
without Higgs violates the tree-level unitarity  at
$\sqrt{s}=1.7\unit{TeV}$ \cite{tBESS,DominiciNuovoCim}.

In our effective Lagrangian, the scattering amplitudes include the exchange
of the new resonances, the scalar one and the vector ones.
At tree-level the amplitudes read
\begin{eqnarray}
  \cM(W_L^+W_L^-\rightarrow W_L^+W_L^-) &=& A(s,t,u)+A(t,s,u),
  \nonumber\\
  \cM(Z_L Z_L\rightarrow Z_L Z_L) &=& 0,
  \nonumber\\
  \sqrt{2}\cM(W_L^+W_L^-\rightarrow Z_L Z_L) &=& A(s,t,u),
  \nonumber\\
  \cM(W_L^\pm Z_L\rightarrow W_L^\pm Z_L) &=& A(t,s,u),
  \nonumber\\
  \sqrt{2}\cM(W_L^\pm W_L^\pm\rightarrow W_L^\pm W_L^\pm) &=& A(t,s,u)+A(u,t,s),
  \nonumber
\end{eqnarray}
where
\begin{eqnarray}
  A(s,t,u) &=& \frac{s}{4v^2}(4-3\alpha)+
        \frac{\alpha M_V^2}{4v^2}[\frac{u-s}{t-M_V^2}+\frac{t-s}{u-M_V^2}]
  \nonumber\\
        && -\frac{a^2}{v^2}\frac{s^2}{s-M_H^2}.
  \label{eq:Astu}
\end{eqnarray}
The widths of both resonances have been neglected;
this is justifiable
as long as their masses are far from the unitarity limit
when measured in terms of the widths.
Recall that $a=1$ corresponds to the SM Higgs boson case
and that $\alpha=(2M_V/g''v)^2$.

\begin{figure}
\centering
\includegraphics[scale=0.75]{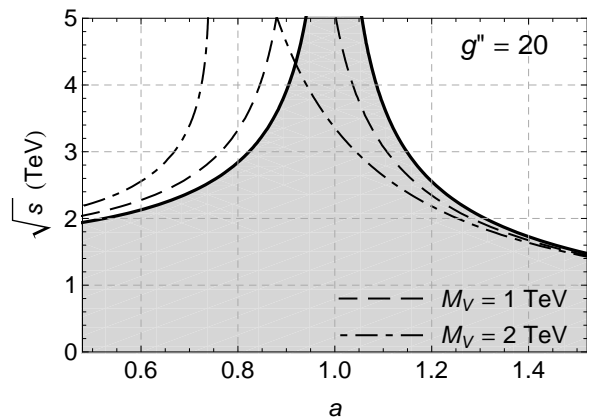}
\caption{\label{fig:Unitarity1}
         The tree-level unitarity constraints from the electroweak
         gauge-boson scattering as functions of $a$. The shaded
         area under the solid line depicts the unitarity allowed
         region for the effective Lagrangian without the vector resonance.
         The dashed and dot-dashed lines indicate the shift of the region
         when the $1\unit{TeV}$ or $2\unit{TeV}$ vector triplets, respectively,
         are added. In both cases $g''=20$.
         Zero decay widths of the new resonances are assumed.
         }
\end{figure}

Obviously, the unitarity constraints obtained from Eqs.~(\ref{eq:Astu})
depend on the mass of the vector resonance as well as on
the degree of anomalousness of the scalar resonance coupling $a$.
Once $a\neq 1$ the Higgs can no longer guarantee unrestricted unitarity.
This problem can be assisted with by invoking additional resonances.
The tree-level unitarity constraints as functions of $a$ obtained for various masses
of the vector triplet and $g''=20$ are shown in
Fig.~\ref{fig:Unitarity1}.
It illustrates the inter-play between the non-SM 125-GeV
scalar resonance and the vector triplet in securing the unitarity of
the gauge boson scattering amplitudes.
With the scalar resonance only, as $a$ departs from one the unitarity limit lowers.
Adding the vector resonance triplet tends to improve
the unitarity for some $a<1$. On the other hand, when $a>1$, it further lowers
the unitarity limit.

\begin{figure}
\centering
\includegraphics[scale=0.75]{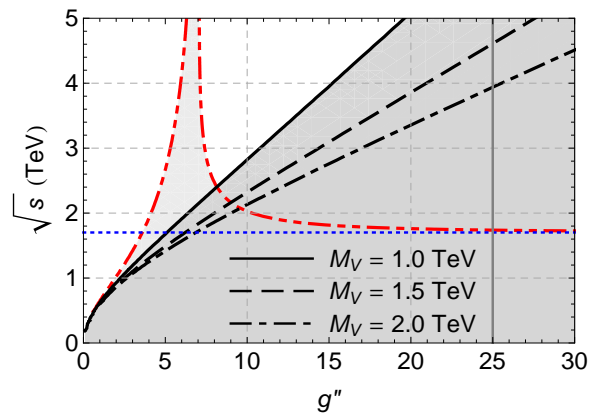}
\caption{\label{fig:Unitarity2}(color online)
         The tree-level unitarity constraints from the electroweak
         gauge-boson scattering as functions of $g''$. The shaded
         areas indicate regions where the unitarity holds.
         The horizontal blue dotted line is the Higgs-less SM unitarity
         limit of 1.7~TeV. The red dot-dot-dashed line is the unitarity
         limit when there is the $1\unit{TeV}$ vector resonance and no Higgs.
         The black lines indicate the unitarity constraints
         for the effective Lagrangian with the $a=1$ scalar and
         the $1\unit{TeV}$ (solid) or $1.5\unit{TeV}$ (dashed)
         or $2\unit{TeV}$ (dot-dashed) vector resonances.
         Zero decay widths of the new resonances are assumed.
         \blue{The vertical gray line at $g''=25$ indicates the position
         of the naive perturbativity limit.}
         }
\end{figure}

\begin{figure}
\centering
\includegraphics[scale=0.75]{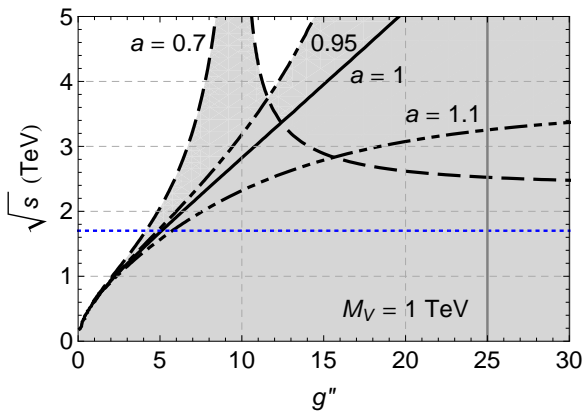}
\caption{\label{fig:Unitarity3}(color online)
         The tree-level unitarity constraints from the electroweak
         gauge-boson scattering as functions of $g''$
         for different values of the scalar to gauge boson coupling:
         $a=1$ (solid), $a=0.7$ (dashed), $a=0.95$ (dot-dashed),
         and $a=1.1$ (dot-dot-dashed).
         The $1\unit{TeV}$ vector resonance and no decay widths are considered.
         }
\end{figure}

Figure~\ref{fig:Unitarity2} illustrates the expected behavior:
while adding the vector resonance to the Higgs-less SM improves its unitarity
limit, it introduces the unitarity problem to the SM with the SM Higgs boson
present.
\blue{Note that in the graph the gray vertical line at $g''=25$ indicates the position
of the naive perturbativity limit. In the same way the limit will be shown in
all following graphs whenever appropriate.}
In Fig.~\ref{fig:Unitarity3} we show the tree-level unitarity restrictions
for the effective Lagrangian when $M_V=1\unit{TeV}$ and
$a$ assumes some non-SM values.

\begin{figure}
\centering
\includegraphics[scale=0.75]{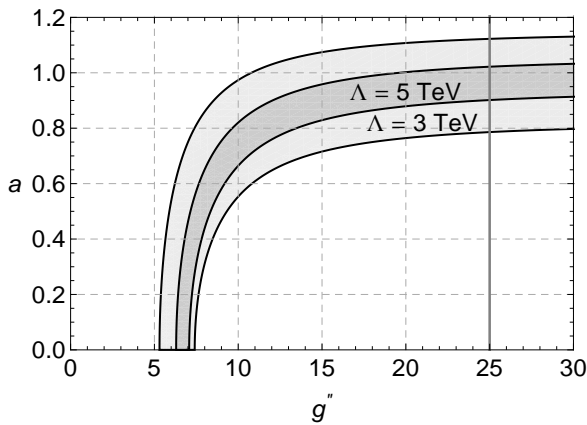}
\caption{\label{fig:Unitarity4}
         The allowed regions of the $a$ and $g''$ parameter space should
         the tree-level unitarity hold up to $3\unit{TeV}$ (light gray) and
         $5\unit{TeV}$ (dark gray).
         The $1\unit{TeV}$ vector resonance and no decay widths are considered.
         }
\end{figure}

Should the unitarity of the effective Lagrangian hold up
to a certain energy the allowed region for the values
of $a$ and $g''$ can be constructed. The allowed region
has a shape of a bent stripe.
In particular, the allowed regions for the unitarity constraints
of $3$ or $5\unit{TeV}$ when $M_V=1\unit{TeV}$ are depicted
in Fig.~\ref{fig:Unitarity4}.

These findings imply that to secure the tree-level unitarity
when $a$ decreases away from its SM value either higher vector resonance
mass or lower $g''$ have to be invoked. The role of the vector resonance
is destructive though if $a$ departs from its SM value in the opposite direction.
Of course, in the approximation we use
these conclusions are independent of the fermion sector structure
of the effective Lagrangian. The observed behavior depends on
properties of scalar and vector resonances only.
Adding the $SU(2)_{L+R}$ triplet axial-vector resonances or
introducing additional constraints on the resonance couplings due to the assumption
that the scalar resonance is a pseudo Nambu--Goldstone boson of some
sort could alter these conclusions.
The investigation of the unitarity question could be refined by considering
the gauge boson decay widths
and/or additional scattering amplitudes. This would make the unitarity
limits sensitive also to the properties of the fermion sector.

%% file: 04a-integrating_out.tex

If there is the tBESS vector resonance triplet
we can learn about its parameters even before its discovery
by measuring deviations
of the known particle couplings from their SM values.
For example, due to the mixing between the vector resonance and
the EW gauge bosons the deviations from the SM values would
be present in the couplings of the EW gauge bosons to the SM fermions.
In this sense, in the case of our effective Lagrangian
the most interesting
vertices should be those of the top and bottom quarks: $Wtb$, $Zbb$,
and $Ztt$.

Unfortunately, the measurements of the $Wtb$ and $Ztt$ vertices
has been rather coarse so far~\cite{TevatronWtb}.
On the other hand, the couplings of the light fermion vertices including
$Zbb$ have  been measured at previous colliders, sometimes to a very high
precision. We refer to these measurement as the low-energy measurements.
While the LHC is capable to refine these measurements, and it has done
so already, the existing improvement are not sufficient to
compete with the low-energy restrictions on the tBESS parameters.
Thus, in our analysis we will focus on the low-energy measurements.

To confront the tBESS free parameters with the low-energy
measurements performed at $\cO(10^2)\unit{GeV}$, we derive
the low-energy (LE) Lagrangian by integrating out the vector resonance
triplet the assumed mass of which is $\cO(10^3)\unit{GeV}$.
It proceeds by taking
the limit $M_{triplet}\rightarrow\infty$, while $g''$ is finite
and fixed, and by substituting the vector resonance
equation of motion (EofM) obtained under these conditions.
The EofM in the unitarity gauge reads
\begin{equation}\label{eq:EofM}
 i\frac{g''}{2}V_\mu^a = \frac{1}{2}(igW_\mu^a+ig'B_\mu\delta^{a3}),
\end{equation}
where $a=1,2,3$.
After the EofM is substituted into the unitary gauge ESB
Lagrangian~(\ref{eq:LagrESBgaugeU}) the alpha multiplied
trace term disappears. We end up with
\begin{eqnarray}\label{eq:LELagrESBgaugeU}
  \Lagr_\mathrm{ESB}^\mathrm{LE} &=&  \frac{1}{2}\pard_\mu h\,\pard^\mu h
                         -\frac{1}{2}M_h^2 h^2
                         -\frac{v^2}{4}\Tr\left(\BW_\mu-\BB_\mu\right)^2
  \nonumber\\
  && \times\,\left(1+2a\frac{h}{v}+a'\frac{h^2}{v^2}+\ldots\right).
\end{eqnarray}
Of course, the EofM has to be substituted in all Lagrangian terms
where the vector resonance field occurs. Then, after the gauge
boson mass matrix diagonalization and the renormalization of the gauge
boson fields the low-energy limit of our effective Lagrangian
is obtained.

Since $\alpha\rightarrow\infty$, the number of free parameters
in the ESB sector has dropped by one. In the low-energy Lagrangian
it is convenient to introduce and use parameters $e$ and $s_\theta$
that are related to the strengths of the charged and neutral currents;
actually, $e$ is the electric charge and $s_\theta$ a counterpart
of the Weinberg angle sine. In addition, the parameter $x$ encodes
the low-energy residues of the new interaction of the vector resonance triplet.
The relations of $e$, $s_\theta$, and $x$ to the parameters of the full
effective Lagrangian are
\begin{equation}
  e = \frac{gg'/G}{\sqrt{1+\left(\frac{gg'}{Gg''/2}\right)^2}},\;\;\;
  s_\theta = g'/G,\;\;\;
  x = g/g'',
\end{equation}
where $G=(g^2+g^{\prime 2})^{1/2}$.
The EW gauge boson masses expressed in terms of $e$, $s_\theta$, and $x$
are given by
\begin{equation}
 M_W^2 = \frac{g_\mathrm{LE}^2\,v^2}{4},\gap
 M_Z^2 = \frac{G_\mathrm{LE}^2\,v^2}{4},
\end{equation}
where $g_\mathrm{LE}$ and $G_\mathrm{LE}$ are the LE strengths of the charged
and neutral currents, respectively. Namely,
\begin{equation}
 g_\mathrm{LE}^2 = \frac{1+4s_\theta^2 x^2}{1+x^2}\,\frac{e^2}{s_\theta^2},
 \;\;\;
 G_\mathrm{LE}^2 = \frac{(1+4s_\theta^2 x^2)^2}{c_\theta^2+x^2}\,\frac{e^2}{s_\theta^2},
\end{equation}
where $c_\theta=(1-s_\theta^2)^{1/2}$.

The EofM also modifies the $\cL_{(t,b)}^\mathrm{tBESS}$ term in~(\ref{eq:tBESSLagrFerm}). Thus
\begin{equation}\label{eq:LagrFermLE}
  \cL_{\mathrm{ferm}}^{\mathrm{LE}} \equiv
  \cL_{\mathrm{ferm}}^{\mathrm{SM}} + \cL_{(t,b)}^{\mathrm{LE-tBESS}} + \cL_\mathrm{ferm}^\mathrm{scalar}.
\end{equation}
In the EW gauge boson mass eigenstate basis and after the proper renormalization
the relevant parts of the $\cL_{\mathrm{ferm}}^{\mathrm{LE}}$ can be expressed as
\begin{eqnarray}\label{eq:LagrFermLE1}
 \cL_{\mathrm{ferm}}^{\mathrm{SM}} + \cL_{(t,b)}^{\mathrm{LE-tBESS}}
  &=& i\bar{\psi}\dslash \psi -e\bar{\psi}\Aslash Q\psi
 \nonumber\\
 &&  -\frac{G_\mathrm{LE}}{2}\bar{\psi}\Zslash (C_LP_L+C_RP_R)\psi
 \nonumber\\
 && - \frac{g_\mathrm{LE}}{\sqrt{2}}\bar{\psi}(\Wslash^+\tau^++\Wslash^-\tau^-)
 \nonumber\\
 &&\phantom{- \frac{g_\mathrm{LE}}{\sqrt{2}}} \times(D_LP_L+D_RP_R)\psi,\;\;\;\;\;
\end{eqnarray}
where $\tau^\pm=\tau^1\pm i\tau^2$, $P_{L,R}=(1\mp\gamma_5)/2$.
For the light fermions (all SM fermions except the top and bottom quarks)
$D_L = 1$, $D_R = 0$, and
\begin{equation}
  C_L = 2T_L^3-2\kappa s_\theta^2 Q,\gap
  C_R = -2\kappa s_\theta^2 Q,
\end{equation}
where
\begin{equation}
  \kappa = \frac{1+2x^2}{1+4s_\theta^2 x^2}.
\end{equation}
In the case of the top and bottom quarks,
\begin{eqnarray}
 C_L &=& 2(1-\Delta L/2)\, T_L^3-2\kappa s_\theta^2 Q,
 \label{eq:CLtopbottomquarks}\\
 C_R &=& 2(P_f\,\Delta R/2)\, T_R^3-2\kappa s_\theta^2 Q,
 \label{eq:CRtopbottomquarks}
\end{eqnarray}
where $P_t = 1$, $P_b = p^2$, and
\begin{equation}\label{eq:DLRtopbottomquarks}
 D_L = 1-\Delta L/2,\gap D _R = p\; \Delta R/2,
\end{equation}
where
\begin{equation}\label{eq:DeltaLR}
 \Delta L = b_L -2\lambda_L,\gap
 \Delta R = b_R +2\lambda_R.
\end{equation}
Hence, the number of free parameters has been reduced also in the fermion
sector of the low-energy Lagrangian. The low-energy observables
will depend on the combinations (\ref{eq:DeltaLR}) of $b$ and $\lambda$
parameters only.
Therefore no limits derived from the low-energy measurements can apply
to $b$'s and $\lambda$'s individually.

In order to make numerical predictions the model under
consideration must
be supplied with an experimental input.
The appropriate experimental input in the case of the LE
Lagrangian consists of the measured value of the Fermi constant $G_F$,
the fine structure constant $\alpha$ at the energy
scale $M_Z$, and the mass $M_Z$ of the $Z$ boson.
It will prove convenient
to replace $G_F$ with the sine of the SM Weinberg angle $s_0$
using the SM relation
\begin{equation}\label{eq:GFinSM}
 \frac{G_F}{\sqrt{2}} = \frac{2\pi\,\alpha(M_Z)}{(2s_0c_0)^2 M_Z^2},
\end{equation}
where $c_{0}= (1-s_0^2)^{1/2}$.
Thus, given the experimental values of $\alpha(M_Z)$, $M_Z$, and $G_F$,
$s_0 $ can be considered as a replacement of $G_F$ in this list.

The value of $e$ is just a synonym of the measured value of $\alpha$,
$e^2=4\pi\alpha$. The non-trivial question is how to properly trade
$s_0$ and $M_Z$ for $s_\theta$ and $v$. For this
we have to write down the LE formula for $G_F$,
\begin{equation}\label{eq:GFinLEtBESS}
   \frac{G_F}{\sqrt{2}}=\frac{2\pi\,\alpha(M_Z)}{(2s_\theta c_\theta)^2 M_Z^2}
                        \frac{(1+4s_\theta^2 x^2)^2}{1+(\frac{x}{c_\theta})^2}.
\end{equation}
Then, comparing~(\ref{eq:GFinLEtBESS}) with (\ref{eq:GFinSM}) we obtain
the implicit relation for $s_\theta(s_0,x)$:
\begin{equation}\label{eq:s0c0vssThetacTheta}
 s_0 c_0 = s_\theta c_\theta \frac{\sqrt{1+(\frac{x}{c_\theta})^2}}{1+4s_\theta^2 x^2}.
\end{equation}
To replace the parameter $v$ the LE formula for the $Z$ boson mass,
\begin{equation}\label{eq:MZinLEtBESS}
  M_Z = \frac{e v}{2s_0 c_0},
\end{equation}
can be considered.

Regarding the remaining parameters
the value of $M_h$ is given by the mass
of the recently discovered candidate of the Higgs boson.
Thus, in the following analysis we will set $M_h= 125\unit{GeV}$.
Parameters fixed by experiment also include the fermion masses.
The existing restrictions on $a,a',\ldots$ and $c_f,c_f',\ldots$
have been discussed at the end of Section~\ref{sec:tBESS}.
As it was also indicated, there we will perform
the best-fit analysis under the simplifying assumption
that the scalar resonance couplings are those of the SM.
It leaves us with four free parameters that will be used
to fit the observables: $x$, $\Delta L$, $\Delta R$, and $p$.
In principle, there is also the cut-off scale $\Lambda$ of
the LE effective Lagrangian. This is usually related to
the mass of the integrated out vector resonance.
Our analysis will be performed with the cut-off scale fixed.

%% file: 04b-LEtBESSpredic.tex

The deviations of the LE Lagrangian from its SM counterpart
modify predictions for the low-energy observables. Thus we can use
their measured values to derive the preferences and restrictions on
the LE free parameters.

In particular,
the experimental limits for the LE-tBESS parameters will be derived
by fitting the low-energy (pseudo-)observables\footnote{
The quantities $\Gamma_b$ and BR$(\bgs)$ are more intimately
related to actual observables than the epsilons. To stress this
fact one might wish to nickname the epsilons as \textit{pseudo-observables}.
Nevertheless, in the following text we will not make this distinction
and will call the epsilons as observables, too.
},
$\epsilon_1$, $\epsilon_2$, $\epsilon_3$,
$\Gamma_b(Z\rightarrow b\bar{b})$, and BR$(\bgs)$.
The epsilons are related to the \textit{basic
observables}~\cite{EpsilonMethod}: the ratio of the electroweak
gauge boson masses, $r_M\equiv M_W/M_Z$; the inclusive partial
decay width of $Z$ to the charged leptons,
$\Gamma_\ell(Z\rightarrow\ell\bar{\ell}+\mathrm{photons})$; the
forward-backward asymmetry of charged leptons at the $Z$-pole,
$A_\ell^{FB}(M_Z)$; and the inclusive partial decay width of $Z$
to bottom quarks, $\Gamma_b(Z\rightarrow b\bar{b}+X)$.

The deviations of $r_M$, $\Gamma_\ell$, and $A_\ell^{FB}$ from
their predicted SM tree level values including the QED and QCD
loop contributions are parameterized by the \textit{dynamical
corrections} $\Delta r_W$, $\Delta\rho$, and $\Delta k$ as
follows~\cite{EpsilonMethod}:
\begin{equation}\label{eq:rM}
  \left(1-r_M^2\right)r_M^2 =
                    \frac{\pi\alpha(M_Z)}{\sqrt{2}G_F M_Z^2 (1-\Delta r_W)}
\end{equation}
and
\begin{eqnarray}
  \Gamma_\ell &=& \frac{G_F M_Z^3}{6\pi\sqrt{2}}
                        (g_A^\ell)^2(1+r_g^2)\left(1+\frac{3\alpha}{4\pi}\right),
  \label{eq:GammaLepton}\\
     A_\ell^{FB} &=& \frac{3r_g^2}{(1+r_g^2)^2},
  \label{eq:AFBlepton}
\end{eqnarray}
where
\begin{equation}
  g_A^\ell = -\frac{1}{2}\left(1+\frac{\Delta\rho}{2}\right),
  \gap
  r_g = \frac{g_V^\ell}{g_A^\ell} = 1-4(1+\Delta k)s_0^2.
\end{equation}

The three epsilons can be defined as the combinations of the
dynamical corrections~\cite{EpsilonMethod}:
\begin{eqnarray}
 \epsilon_1 &=& \Delta\rho,
 \label{eq:eps1toDynamCorrs}\\
 \epsilon_2 &=& c_0^2\Delta\rho + \frac{s_0^2}{c_{20}}\Delta r_W - 2s_0^2\Delta k,
 \label{eq:eps2toDynamCorrs}\\
 \epsilon_3 &=& c_0^2\Delta\rho + c_{20}\Delta k,
 \label{eq:eps3toDynamCorrs}
\end{eqnarray}
where $s_0$ ($c_0$) was defined in Eq.~(\ref{eq:GFinSM}),
and $c_{20}\equiv c_0^2-s_0^2$.

The $Zbb$ vertex is naturally tested in the $Z\rightarrow
b\bar{b}+X$ decay. The corresponding decay width formula
reads~\cite{EpsilonMethod}
\begin{eqnarray}\label{eq:GammaBottom}
  \Gamma_b &=& \frac{G_F M_Z^3}{6 \pi \sqrt{2}} \beta
               \left[\frac{3-\beta^2}{2}(g_{V}^b)^2 + \beta^2 (g_{A}^b)^2\right] \times
  \nonumber\\
           &&  N_C R_{\mathrm{QCD}} \left(1+ \frac{\alpha}{12\pi} \right),
\end{eqnarray}
where $\beta = (1-4m_b^2/M_Z^2)^{1/2}$, and $R_{\mathrm{QCD}} =
1+1.2 a - 1.1a^2 -13a^3$ is the QCD correction factor, $a =
\alpha_s(M_Z)/\pi$.

The precise measurement of $\Gamma_b$ can uncover whether the
bottom quark anomalous couplings $g_{V,A}^b$ differ from the
anomalous couplings of other charged SM fermions. Assuming the
couplings differ in their $SU(2)_L$ parts only, the standard
parameterization of the difference is by introducing the parameter
$\epsilon_b$~\cite{EpsilonMethod}:
\begin{eqnarray}
  g_A^b &=& g_A^\ell (1+\epsilon_b),
  \label{eq:gABfromDynamicalCorrs}\\
  g_V^b &=& \left(1+\frac{\Delta\rho}{2}\right)
                    \left[-\frac{1}{2}(1+\epsilon_b)+\frac{2}{3}(1+\Delta k)s_0^2\right].
  \label{eq:gVBfromDynamicalCorrs}
\end{eqnarray}

However, our effective Lagrangian admits a more general
pattern of the bottom versus light quark anomalous coupling
difference than it is assumed in the definition of $\epsilon_b$.
In our effective Lagrangian the $\epsilon_b$ definition assumptions are met
when either $p=0$, or $b_R=-2\lambda_R$. Otherwise, the
experimental value of $\Gamma_b$ rather then
$\epsilon_b^{\mathrm{exp}}$ must be related to
theoretical prediction in order to derive the low-energy limits on
the tBESS free parameters.

The scalar resonance couplings
do not contribute to the dynamical corrections at tree level.
Thus, the tree-level contributions of the LE Lagrangian to
the $\Delta r_W$, $\Delta\rho$, and $\Delta k$
as well as to $g_{V,A}^b$ are given
by the vector resonance sector only.
They read
\begin{equation}\label{eq:DeltaRHOandKfromLEtBESS}
  \Delta\rho = 0,\gap
  \Delta k = \left(\frac{s_\theta}{s_0}\right)^2
  \kappa(x,s_\theta)\,-\,1,
\end{equation}
\begin{eqnarray}
 g_V^b &=&
 \frac{1}{4}(\Delta L - p^2\;\Delta R)
 +\frac{2}{3}\Delta k s_0^2,
 \label{eq:gVBfromLEtBESS}\\
 g_A^b &=&
 \frac{1}{4}(\Delta L + p^2\;\Delta R).
 \label{eq:gABfromLEtBESS}
\end{eqnarray}
The tree-level contribution to $\Delta r_W$ is obtained from
the LE expression for the ratio $r_M = M_W/M_Z$. It reads
\begin{equation}
  r_M^2 = \frac{c_\theta^2+x^2}{(1+4s_\theta^2 x^2)(1+x^2)}.
\end{equation}
Then
\begin{equation}\label{eq:DeltaRWfromLEtBESS}
 \Delta r_W = 1-\left(\frac{1+x^2}{1+2x^2}\right)^2.
\end{equation}

Now, the tree-level contributions to $\epsilon_i$'s can be
obtained from the dynamical corrections using
Eqs.~(\ref{eq:eps1toDynamCorrs}) through
(\ref{eq:eps3toDynamCorrs}):
\begin{eqnarray}
 \epsilon_{1} &=& 0,
 \nonumber\\
 \epsilon_{2} &=&
 \frac{s_0^2}{c_{20}}\frac{x^2(2+3x^2)}{(1+2x^2)^2}-2s_0^2\,\Delta k(x),
 \nonumber\\
 \epsilon_{3} &=& c_{20}\,\Delta k(x),
 \nonumber
\end{eqnarray}
where $\Delta k(x)$ is given in (\ref{eq:DeltaRHOandKfromLEtBESS}).
The leading terms of the epsilon expansions in powers of $x$ at $x=0$ read
\begin{eqnarray}
 \epsilon_{2} &=& -2.71 \,x^4 + 2.96 \,x^6 + \ldots,
 \label{eq:eps2LE0powerseries}\\
 \epsilon_{3} &=& x^2 + 0.66 \,x^4 + 2.56 \,x^6 + \ldots.
 \label{eq:eps3LE0powerseries}
\end{eqnarray}

There is no reason to expect that the LE anomalies at the
tree level overwhelm the 1-loop contributions of the LE
Lagrangian to the epsilons. Thus, both contributions should be
considered when predicting the epsilon observables
\begin{equation}
 \epsilon_i = \epsilon_i^{\mathrm{LE(0)}}
                          + \epsilon_i^{\mathrm{LE(1)}},
 \gap i=1,2,3,b,
\end{equation}
where $\mathrm{LE(0)}$ and $\mathrm{LE(1)}$ denotes
the tree-level and 1-loop contributions of the
LE Lagrangian, respectively.

Since we study an effective non-renormalizable Lagrangian, it is
not that obvious how to properly deal with the higher order
calculations~\cite{Malkawi}. One does not know the underlying
theory therefore there is no way to establish a correct scheme for
the effective Lagrangian~\cite{Georgi}. While the divergent piece
in loop calculations can be associated with a physical cut-off
$\Lambda$ up to which the effective Lagrangian is
valid~\cite{PecceiZhang}, for the finite piece there is no
completely satisfactory approach available~\cite{BurgessLondon}.

We approximate $\epsilon_i^\mathrm{LE(1)}$ by the sum of the SM weak loop
corrections $\epsilon_i^\mathrm{SM(1)}$ representing the loop contributions
from the scalar resonance and the vector resonance related loop contributions
$\epsilon_i^{\mathrm{vec}(1)}$:
\begin{equation}\label{eq:EpsilonLoopApproxHiggsless}
  \epsilon_i^{\mathrm{LE}(1)} \approx \epsilon_i^{\mathrm{SM(1)}}
                   + \epsilon_i^{\mathrm{vec}(1)}.
\end{equation}

The $\epsilon_i^{\mathrm{SM(1)}}$ contributions are given by
the following relations~\cite{newBESS}:
\begin{eqnarray}
  \epsilon_1^{\mathrm{SM(1)}} &=& \left(+5.60-0.86\ln\frac{M_h}{M_Z}\right)\times 10^{-3},
  \label{eq:eps1SM}\\
  \epsilon_2^{\mathrm{SM(1)}} &=& \left(-7.09+0.16\ln\frac{M_h}{M_Z}\right)\times 10^{-3},
  \label{eq:eps2SM}\\
  \epsilon_3^{\mathrm{SM(1)}} &=& \left(+5.25+0.54\ln\frac{M_h}{M_Z}\right)\times 10^{-3},
  \label{eq:eps3SM}\\
  \epsilon_b^{\mathrm{SM(1)}} &=&       -6.43\times 10^{-3}.
  \label{eq:epsbSM}
\end{eqnarray}
For $M_h=125\unit{GeV}$ and $M_Z=91.1876\unit{GeV}$
the following SM contributions are obtained:
$\epsilon_1^{\mathrm{SM(1)}} = 5.33\times 10^{-3}$,
$\epsilon_2^{\mathrm{SM(1)}} = -7.04\times 10^{-3}$, and
$\epsilon_3^{\mathrm{SM(1)}} = 5.42\times 10^{-3}$.

The 1-loop SM contributions to $g_V^b$ and $g_A^b$ can be obtained
by subtracting the SM tree-level couplings from the SM tree plus
1-loop couplings:
\begin{equation}
 (g_{V,A}^b)^{\mathrm{SM(1)}} =
 (g_{V,A}^b)^{\mathrm{SM(0+1)}}-
 (g_{V,A}^b)^{\mathrm{SM(0)}},
\end{equation}
where $(g_{V,A}^b)^{\mathrm{SM(0+1)}}$ are given by
Eqs.~(\ref{eq:gABfromDynamicalCorrs}) and
(\ref{eq:gVBfromDynamicalCorrs}) if $\Delta\rho=(\Delta
\rho)^{\mathrm{SM(1)}}$, $\Delta k=(\Delta k)^{\mathrm{SM(1)}}$,
and $\epsilon_b=\epsilon_b^{\mathrm{SM(1)}}$ are applied. Of
course, $(g_{V}^b)^{\mathrm{SM(0)}}=-1/2+2s_0^2/3$ and
$(g_{A}^b)^{\mathrm{SM(0)}}=-1/2$.

The $\epsilon_i^{\mathrm{vec}(1)}$ contributions can be calculated using
the results of~\cite{LariosKappaAnalysis}. The paper provides
expressions for new physics loop contributions to the epsilon
parameters in terms of generic anomalous couplings of the non-linear
electroweak chiral Lagrangian. Up to the order of
$m_t^2\ln\Lambda^2$ the anomalous loop contributions
read
\begin{eqnarray}
  \epsilon_1^{\mathrm{NP(1)}} &=& \frac{3 m_t^2 G_F}{2 \sqrt{2} \pi^2}\ln\frac{\Lambda^2}{m_t^2}
                       \left[ \kappa_L^{Wtb}\left( 1+{\kappa_L^{Wtb}}\right)\right.
  \nonumber\\
                   & & +
  \left. \left(\kappa_R^{Ztt}-\kappa_L^{Ztt}\right)\left(1-\kappa_R^{Ztt}+\kappa_L^{Ztt}\right) \right],
  \label{eq:LariosEps1}\\
  \epsilon_2^{\mathrm{NP(1)}} &=& \epsilon_3^{\mathrm{NP(1)}} \;=\; 0\,,
\end{eqnarray}
where $\Lambda$ is the cut-off scale of the effective Lagrangian
under consideration. In the cases when the $\mathrm{NP(1)}$
contributions depend on $\kappa^{Zbb}$ the dependence is
suppressed by $m_b\ll m_t$.

In our case, when $\Lambda=1$~TeV and using the numerical values of
\ref{app:ExpValues}, the leading terms of the $x^2$ series
of $\epsilon_1^{\mathrm{vec}(1)}$ read
\begin{eqnarray}
 \frac{\epsilon_1^{\mathrm{vec}(1)}}{10^{-2}} &=& \phantom{-}6.57 \,\Delta R
 - 2.82 \,(2 - 3\,\Delta L) \, x^2 + \ldots,
 \label{eq:eps1dLE1series}
\end{eqnarray}
where we have also neglected non-linear terms in $\Delta L$,
$\Delta R$. The $\epsilon_1^{\mathrm{vec}(1)}$ series
for $\Lambda=2$~TeV is obtained when (\ref{eq:eps1dLE1series})
is multiplied by the numerical factor of $1.39$.

In our analysis, we have not calculated $(g_{V,A}^b)^{\mathrm{vec}(1)}$.
The fit is based on the $\mathrm{LE(0)}$ and
$\mathrm{SM(1)}$ contributions to $g_{V,A}^b$ only.
Thus obtained $g_{V,A}^b$ are, in turn, substituted to
Eq.~(\ref{eq:GammaBottom}).
We justified this simplifying approximation by comparing
the single-observable fits based on
$\epsilon_b^{\mathrm{LE(0)}+\mathrm{SM(1)}+\mathrm{vec}(1)}$ with the
fits based on $(g_{V,A}^b)^{\mathrm{LE(0)}+\mathrm{SM(1)}}$ when
$p=0$, see \cite{tBESS}. Figure of \cite{tBESS} illustrates that
the absence of the $\mathrm{vec}(1)$ contribution in the latter fits
introduces only relatively small shifts in the obtained confidence
level contours.

The $B\rightarrow X_s\gamma$ decay puts limits on the anomalous
$W^\pm t_{L}b_{L}$ and $W^\pm t_{R}b_{R}$
vertices~\cite{Malkawi,LariosKappaAnalysis}. In the SM it proceeds
through the flavor changing neutral current loop process
$b\rightarrow s\gamma$ dominated by the top quark exchange
diagram. The $B\rightarrow X_s\gamma$ branching fraction can be
sensitive to physics beyond the SM via new particles entering the
penguin loop. When expressed in terms of the real anomalous $Wtb$
couplings, $\kappa_L^{Wtb}$ and $\kappa_R^{Wtb}$, it can be
approximated by the following formula~\cite{LariosKappaAnalysis}:
\begin{eqnarray}
 \mbox{BR}(\bgs) \times 10^{4} &=&
 3.07 + 280\,\kappa_R^{Wtb} + 2\,\kappa_L^{Wtb}
 \nonumber\\
 & &+  5520\,(\kappa_R^{Wtb})^2 + 0.3\,(\kappa_L^{Wtb})^2
 \nonumber\\
 & &+ 79\,\kappa_L^{Wtb}\kappa_R^{Wtb}.
 \label{eq:BRb2gs}
\end{eqnarray}

The expressions for
the LE anomalous couplings that are needed in the formulas
(\ref{eq:LariosEps1}) and (\ref{eq:BRb2gs})
are given in \ref{app:AnomCplngs}.

%% file: 04c-fits_and_limits.tex

Using the formulas for the LE predictions
we have performed a multi-parameter $\chi^2$ fit of the
observables in order to obtain the most preferred values
and confidence level intervals for the LE-tBESS parameters.
The set of fitted observables consists of
$\epsilon_1$,$\epsilon_2$,$\epsilon_3$,$\Gamma_b$, and $\mbox{BR}(\bgs)$.
The experimental values of the observables
used in this analysis are
shown in \ref{app:ExpValues}.

By fitting the five observables mentioned above
with the four free parameters ---
$x$, $\Delta L$, $\Delta R$, and $p$ ---
we found the best values
\begin{equation}\label{eq:BestFit1}
  g''(x)=29, \quad \Delta L=-0.004, \quad p\;\Delta R=0.003,
\end{equation}
with $\chi^2_{min}=2.40$. Since $\mathrm{d.o.f.}=5-4=1$,
the obtained value of $\chi^2_{min}$ corresponds to the backing
of $12\%$.
Within the rounding errors these values hold for the cut-off
scale between $0.3\unit{TeV}\leq\Lambda\leq 10^3\unit{TeV}$, at least.
The best values of $p$ and $\Delta R$ depend on $\Lambda$, separately;
in particular,
\begin{eqnarray}
  \Lambda=1\unit{TeV}: & \quad \Delta R=0.016 \quad & p=0.209
  \label{eq:bestvaluesL1} \\
  \Lambda=2\unit{TeV}: & \quad \Delta R=0.011 \quad & p=0.289
  \label{eq:bestvaluesL2}
\end{eqnarray}
Note that the best-fit value of $g''$ falls into the regions allowed
by the unitarity. 
Further, the preferred value of $p$
supports the idea about stronger right-top than right-bottom coupling
of the vector resonance.
\blue{The higher value of $g''$ is
certainly preferable considering the assumed strong nature of 
underlying fundamental theory. On the other hand, the best value is somewhat above 
the naive perturbativity limit of 25. Of course, the results of our analysis
are certainly not reliable above perturbativity limit. 
Nevertheless, following the reasons discussed in Section~\ref{sec:tBESS}
we will show results for $g''$ up to 30.}

The observables $\epsilon_1$ and $\Gamma(\bgs)$ are essential in
the explanation the observed behavior of the preferred values of $p$ and $\Delta R$.
The parameters $p$ and $\Delta R$ enter three of the considered observables only.
Namely, these are $\epsilon_1$, $\Gamma_b(Z\rightarrow b\bar{b})$, and $\Gamma(\bgs)$.
While $\epsilon_1$ depends on $\Delta R$ solely, $\Gamma(\bgs)$ depends on the
product $p\,\Delta R$ and $\Gamma_b$ depends on $p^2\Delta R$.
The sensitivity to $\Lambda$ enters through $\epsilon_1$ only.

Since the free parameter space is four-dimensional, it is impossible to
graphically depict the CL regions around the best-fitting point.
Nevertheless, we can quote the marginalized intervals
--- the one-dimensional projections of the confidence region
--- for each parameter. In our case, the marginalized intervals of
the $95\%$ CL region read\footnote{\label{fn1}
The restriction on $p\,\Delta R$ derived from the LHC measurement of
$Wtb$ vertex~\cite{WtbATLAS} reads $-0.40\leq p\,\Delta R\leq 0.46$ when assuming
$\Delta L=0$.}
\begin{equation}
  \begin{array}{ccccc}
    12 & \leq & g'' &  & \\
    -0.013 & \leq & \Delta L & \leq & 0.006 \\
    -0.006 & \leq & \Delta R & \leq & 0.056
  \end{array}
\end{equation}
when $\Lambda = 1\unit{TeV}$ and $p\geq 0$ is assumed.
The value of $p$ is restricted only very mildly by the considered
data. The whole physically motivated interval $0\leq p \leq 1$
belongs to the marginalized interval of the $95\%$ CL region.
The change of the cut-off scale to $\Lambda = 2\unit{TeV}$ alters
only the marginalized interval for $\Delta R$,
\begin{equation}
  \begin{array}{ccccc}
    -0.005 & \leq & \Delta R & \leq & 0.041
  \end{array}
\end{equation}
and leaves all other conclusions basically intact.
Let us note that the limit $g''\geq 12$ is implied by
the restriction $0\leq x\leq 0.056$.

\begin{figure}[h]
\centering
\includegraphics[scale=0.7]{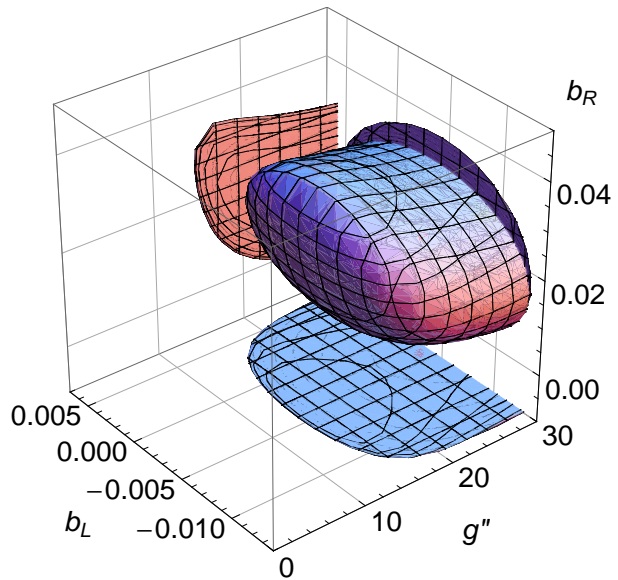}
\caption{\label{fig:3Dcut}(color online)
         The $p=0.209$ cut of the 95\% CL allowed region of
         the parameters $\{g'',\Delta L,\Delta R, p\}$
         when $\Lambda=1\unit{TeV}$.
         The 2D projections of the allowed region
         to $(g'',\Delta L)$, $(g'',\Delta R)$, and
         $(\Delta L, \Delta R)$ planes are also shown.
         }
\end{figure}

Since the value of $p$ can be motivated by theory, e.g.\
by the models of the partial compositeness, it might be useful
to depict a three-dimensional cut of the four-dimensional
$95\%$ CL region for some fixed value of $p$. We choose
the best value of $p=0.209$ for $\Lambda = 1\unit{TeV}$, see
Eq.~(\ref{eq:bestvaluesL1}). The obtained three-dimensional
allowed region of the parameters $x$, $\Delta L$, and $\Delta R$
is shown in Fig.~\ref{fig:3Dcut}.
Only the \blue{$g''\leq 30$} part of the region is shown in the graph.
The region is not restricted in $g''$ from above.

\begin{figure*}
\includegraphics[scale=0.8]{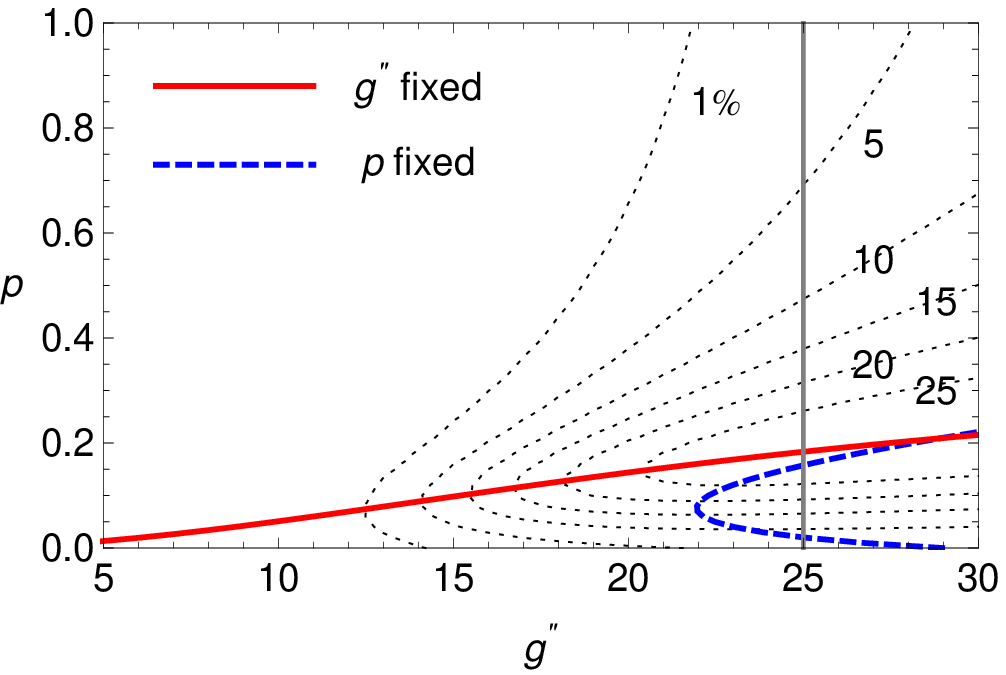}
\gap
\includegraphics[scale=0.8]{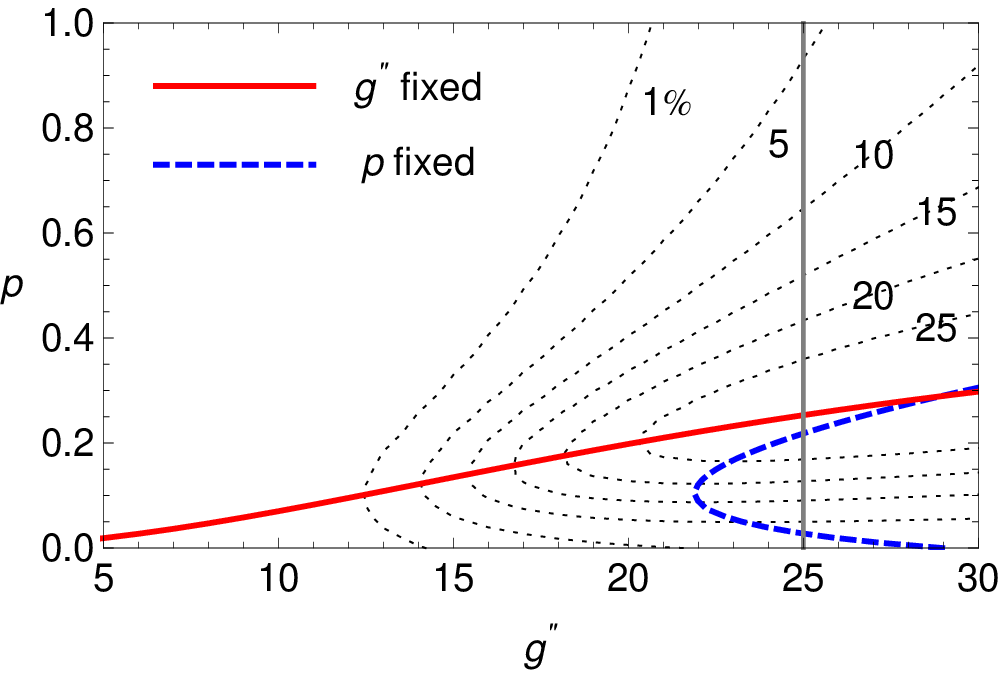}
\caption{\label{fig:fixed_x_p}(color online)
         The graphs of the best-fit values of $p$ (red solid line)
         and $g''$ (blue dashed line) as functions of beforehand fixed values
         of, respectively, $g''$ and $p$. The remaining fitting parameters
         are $\Delta L$ and $\Delta R$.
         The gray dotted contours intersect the curves at points
         with $1\%$, $5\%$, $10\%$, $15\%$, $20\%$, and $25\%$
         backings ($\mathrm{d.o.f.}=2$).
         The intersection of the dashed blue and solid red curves
         possesses a $30\%$ backing.
         The left and right panels correspond to $\Lambda=1\unit{TeV}$
         and $\Lambda=2\unit{TeV}$, respectively.
  }
\end{figure*}

It is certainly interesting to see how the best-fitting values of the free parameters
would change if some of the parameters are fixed, presumably by theoretical assumptions.
\blue{Actually, even if there were no theoretical presumption for fixing 
the value of a particular
parameter by studying plots where some of the free parameters are fixed ahead 
better understanding of behavior of the full four-parameter fit can be achieved.}
One just has to be careful when interpreting such graphs;
e.g., in assigning a correct backing to a set of parameter values.
Having said that, in Fig.~\ref{fig:fixed_x_p} we show the best-fit values
of $x$ ($p$) when the values of $p$ ($x$) have been fixed beforehand.
Of course, $\Delta L$ and $\Delta R$ are the remaining free parameters
in the fit. Thus, $\mathrm{d.o.f.}=5-3=2$ in this case.

\begin{figure}
\includegraphics[scale=0.8]{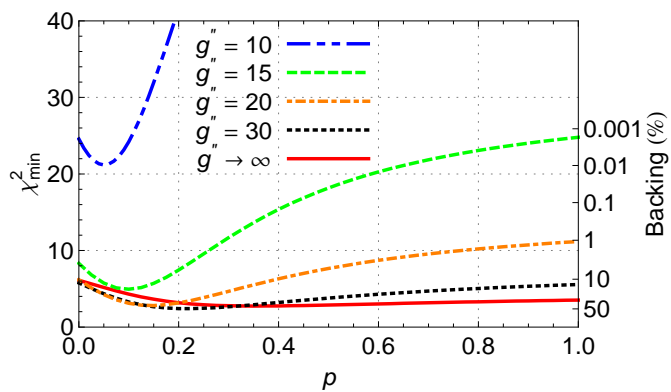}
\caption{\label{fig:chi2min_gpp-p_fixed}(color online)
         $\chi^2_{min}$ of the fit by $\Delta L$ and $\Delta R$ as a function of
         the fixed parameters $g''$ and $p$ for $\Lambda=1\unit{TeV}$.
         The labels on the r.h.s.\ axis indicate the backings for $\mathrm{d.o.f.}=3$.
 }
\end{figure}

The contour dashed lines in Fig.~\ref{fig:fixed_x_p} connect the points
with the same backings in the fits by free parameters $\Delta L$ and $\Delta R$
if both --- $x$ and $p$ --- are fixed beforehand. In this case, $\mathrm{d.o.f.}=5-2=3$.
The $\chi^2_\mathrm{min}$ values for various combinations of fixed $g''$ and $p$
are shown in Fig.~\ref{fig:chi2min_gpp-p_fixed}.
We can see that the best backing for the fits is getting less
pronounced as $g''$ approaches $30$ from below. More specifically,
while backings of the fits with different $p$'s can differ by
several orders of magnitude when $g''\stackrel{<}{\sim} 20$, the
backing for $g''=30$ changes between $10$ and $50\%$ as $p$
crawls along the $\langle 0;1\rangle$ interval.

\begin{figure*}
\centering
\includegraphics[scale=0.75]{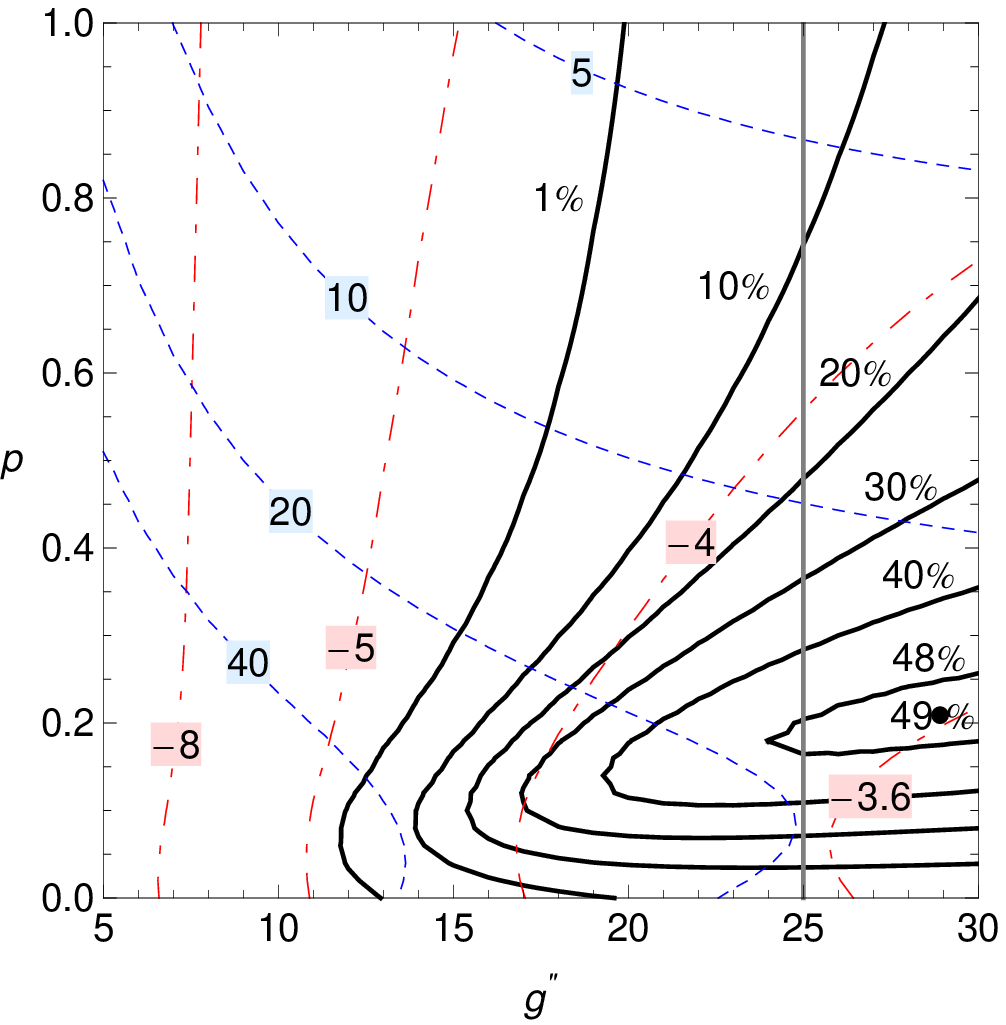}
\hspace{0.5cm}
\includegraphics[scale=0.75]{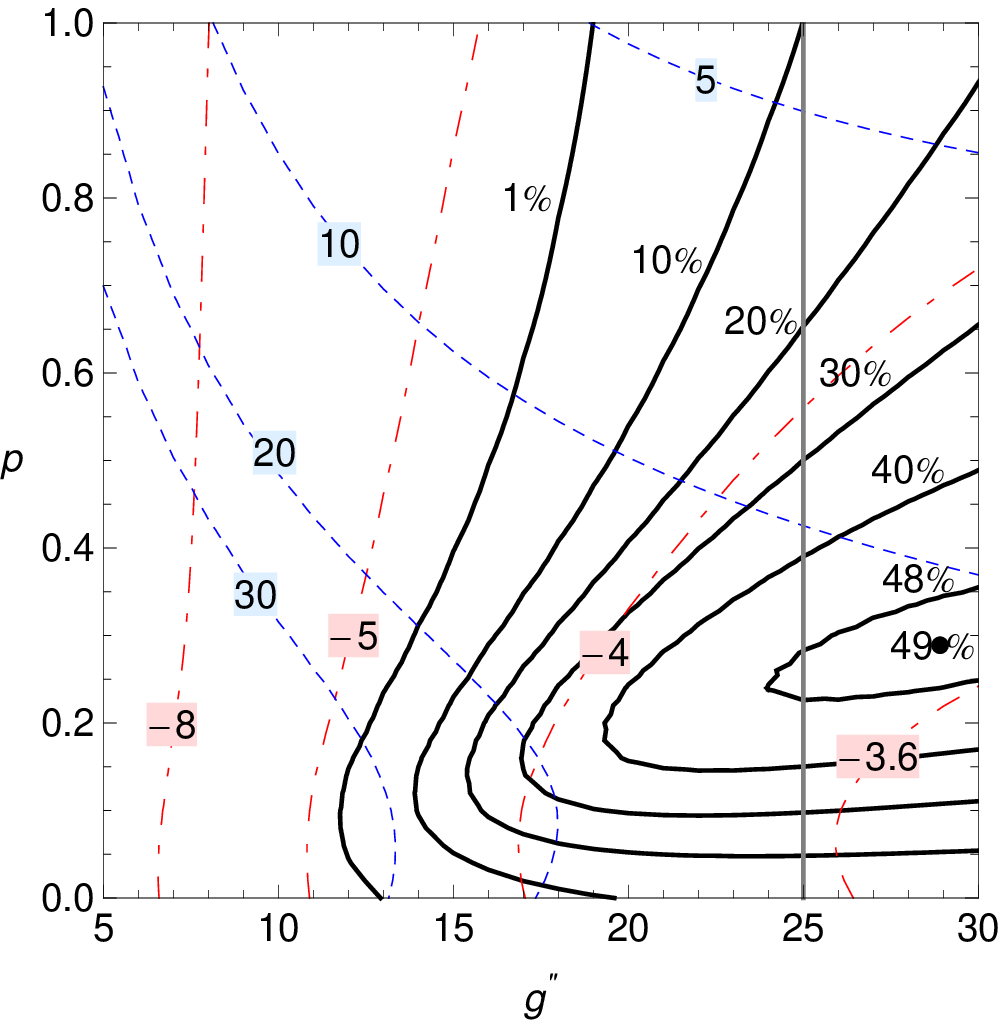}
\caption{\label{fig:MeteoMaps}(color online)
         The contours (black solid lines)
         connecting the $(g'',p)$ points with the same backings
         in the fit by free parameters $\Delta L$ and $\Delta R$.
         The backing values shown on the contours correspond to $\mathrm{d.o.f.} = 3$.
         The graphs also contain the grid from which the best-fit values
         of $\Delta L$ (red dot-dashed) and $\Delta R$ (blue dashed)
         for each given pair of fixed values $(g'',p)$ can be read off.
         The numbers attached to the grid lines are $10^3$ times
         the actual values represented by the lines.
         The left and right panels display the $\Lambda=1\unit{TeV}$
         and $\Lambda=2\unit{TeV}$ cases, respectively.
         }
\end{figure*}

The best-fit values of $\Delta L$ and $\Delta R$ for the given $g''$ and $p$
can be read off of the graphs in Fig.~\ref{fig:MeteoMaps}.
Besides, the contours connecting the $(g'',p)$
points with the same backings are shown there, too.
Table~\ref{tab:fits} lists explicitly some of the best-fit values of $\Delta L$
and $\Delta R$ corresponding to the assortment of fixed $g''$ and $p$ values.

\begin{table}[h]
\centering
\caption{\label{tab:fits}
         The best-fit values of $\Delta L$ and $\Delta R$
         found in the fits for various fixed values of $g''$ and $p$
         when $\Lambda=1$~TeV.
         The backing shown corresponds to $\mathrm{d.o.f}=5-2=3$.
        }
\begin{tabular}{clccrr}
\hline\noalign{\smallskip}
  \multicolumn{2}{c}{fixed} & \multicolumn{2}{c}{best fits} & &\\
  $g''$ & $p$ & $\Delta L$ & $\Delta R$ & $\chi^2_{min}$ & Backing (\%) \\
\noalign{\smallskip}\hline\noalign{\smallskip}
  15 & 0.10   & -0.0042 & 0.0336 & 4.94 & 17.6\phantom{0000} \\
  20 & 0      & -0.0038 & 0.0229 & 6.18 & 10.3\phantom{0000} \\
  20 & 0.14   & -0.0038 & 0.0231 & 2.81 & 42.2\phantom{0000} \\
  20 & 0.5    & -0.0042 & 0.0101 & 7.65 &  5.4\phantom{0000} \\
  20 & 1      & -0.0045 & 0.0046 & 11.77 & 1.1\phantom{0000} \\
  25 & 0.18   & -0.0037 & 0.0181 & 2.44 & 48.6\phantom{0000} \\
  30 & 0.2    & -0.0036 & 0.0157 & 2.42 & 49.1\phantom{0000} \\
\noalign{\smallskip}\hline
\end{tabular}
\end{table}

\begin{figure*}
\centering
\includegraphics[scale=0.70]{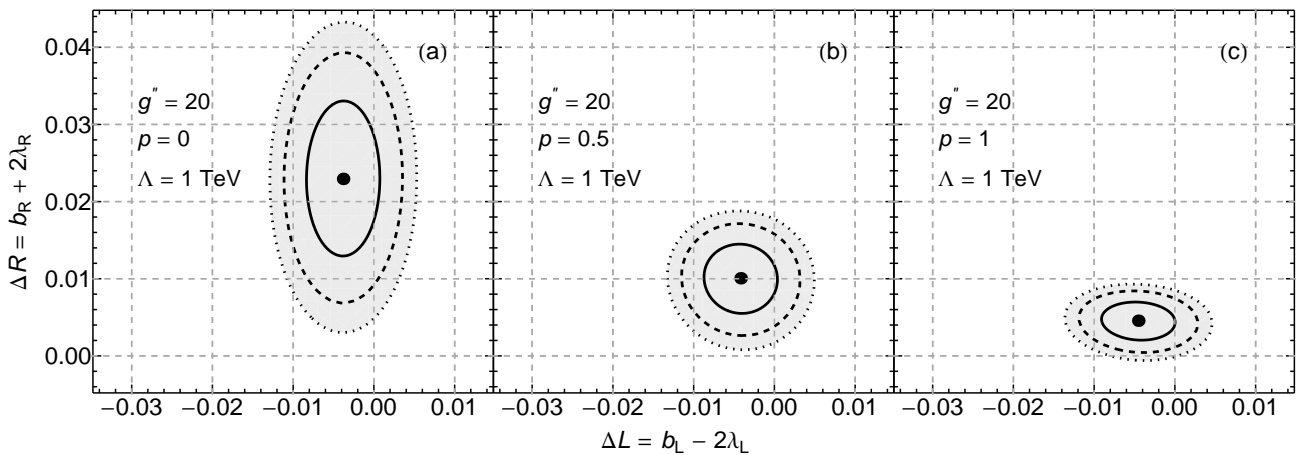}
\caption{\label{fig:CLcontours1}
The 90\% C.L. (solid line), 95\% C.L. (dashed line),
 and 99\% C.L. (dotted line) allowed regions in
 the $(\Delta L, \Delta R)$ parameter space.
 \blue{
 The regions are derived from the two-parameter fit
 by $\Delta L$ and $\Delta R$ for
 $g''=20$, $\Lambda=1$~TeV, and (a) $p=0$,
 (b) $0.5$, and (c) $1$.
 }
 The best-fit values of $\Delta L$ and $\Delta R$
 are indicated by the dots.
 }
\end{figure*}

Eventually, in Fig.~\ref{fig:CLcontours1} we show the allowed CL regions
for the fits around some selected combinations of $g''$ and $p$ parameters
considered also in Table~\ref{tab:fits}.
\blue{In particular, we display contours for $g''=20$ and 
three different values of $p$. Note that
the backings for different values of $p\in \langle 0;1\rangle$ when $g''\approx 20$
do not change significantly (see Figs.~\ref{fig:chi2min_gpp-p_fixed}
and \ref{fig:MeteoMaps}). The backings for $g''=20$, $\Lambda=1\unit{TeV}$, 
and $p=0$, $0.5$, and $1$ are about $10\%$, $5\%$, and $1\%$, respectively.}
Since $\Delta L$ is
predominantly related to a different observable ($\Gamma_b$) than
$\Delta R$ and $p$ ($\epsilon_1$, $\Gamma(\bgs)$), it is not
unexpected that the changes in $p$ affect $\Delta R$ only. In all
three displayed cases the $95\%$~C.L.\ allowed interval for $\Delta
L$ reads \blue{$(-0.012,0.003)$}. The $95\%$~C.L.\ allowed interval for
$\Delta R$ is \blue{$(0.007,0.039)$} when $p=0$. It shrinks to \blue{$\Delta
R\in(0.000,0.008)$} when $p=1$.

\begin{figure*}
\centering
\includegraphics[scale=0.70]{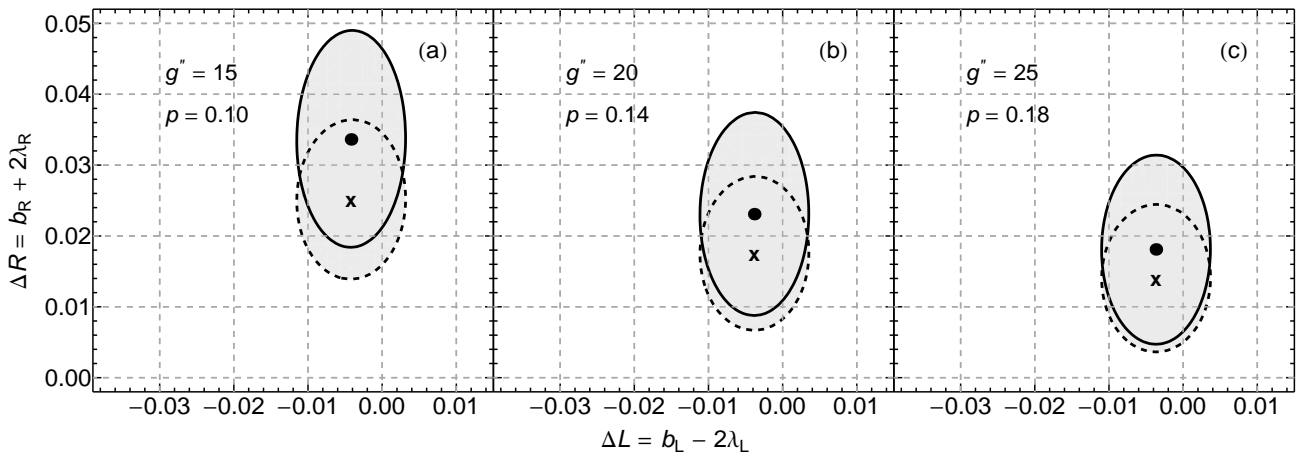}
\caption{\label{fig:CLcontours2}
 The 95\% C.L.\ allowed regions in
 the $(\Delta L, \Delta R)$ parameter space derived from the same
 fit as in Fig.~\ref{fig:CLcontours1} except for
 the values of the fixed parameters.
 The solid contours correspond to $\Lambda=1\unit{TeV}$,
 the dashed ones to $\Lambda=2\unit{TeV}$.
 \blue{
 The values of the fixed parameters $g''$ and $p$ are, respectively,
 (a) $15$ and $0.10$, (b) $20$ and $0.14$, and
 (c) $25$ and $0.18$.}
 The best-fit values of $\Delta L$ and $\Delta R$ are
 indicated by the dot and cross
 for $\Lambda=1\unit{TeV}$ and $\Lambda=2\unit{TeV}$, respectively.
 }
\end{figure*}

The effect of altering $g''$ and/or $\Lambda$ is illustrated
in Fig.~\ref{fig:CLcontours2}. There, the 95\% C.L.\
allowed regions in the $(\Delta L, \Delta R)$ parameter space when
$p$ assumes the values \blue{$0.10$, $0.14$, and $0.18$} are shown.
These are the $p$ values with the highest backings for \blue{$g''=15$, $20$,
and $25$}, respectively. The corresponding best-fit $\Delta L$
and $\Delta R$, $\chi^2_\mathrm{min}$'s and backings are shown in
Table~\ref{tab:fits}. The allowed regions are shown
for the cut-off scales $\Lambda=1\unit{TeV}$ and
$\Lambda=2\unit{TeV}$. As far as the allowed regions are
concerned, $\Delta L$ falls within $(-0.012, 0.004)$,
independently of the other parameter values. For
$\Lambda=1\unit{TeV}$, the $\Delta R$ limits read $(0.018, 0.049)$
when $g''=15$, $(0.009, 0.037)$ when $g''=20$, and $(0.005, 0.032)$
when $g''=25$. For
$\Lambda=2\unit{TeV}$, the $\Delta R$ limits read $(0.014, 0.036)$
when $g''=15$, $(0.007, 0.028)$ when $g''=20$, and $(0.004, 0.024)$
when $g''=25$.

 \blue{
Having the data preferred values of the LE Lagrangian parameters
we can evaluate the
strengths of the direct interaction vertices
of the vector resonance triplet with the top and bottom quarks.
For the sake,
we will recall the vertices of the vector resonance triplet with the top and bottom quarks
that have been derived in~\cite{tBESS}. In the gauge boson flavor basis they read
\begin{eqnarray}
  c^{Vtt}_L = \sqrt{2}\,c^{Vtb}_L = - c^{Vbb}_L &=& -\phantom{p^2}\,b_L\, g''/4
  \nonumber\\
  c^{Vtt}_R \phantom{ = \sqrt{2}\,c^{Vtb}_L = - c^{Vbb}_L } &=& -\phantom{p^2}\,b_R\, g''/4
  \nonumber\\
  \phantom{c^{Vtt}_R =} \sqrt{2}\,c^{Vtb}_R \phantom{= - c^{Vbb}_L } &=& -p\phantom{^2}\, b_R\, g''/4
  \nonumber\\
  \phantom{c^{Vtt}_R = \sqrt{2}\,c^{Vtb}_R =} - c^{Vbb}_R &=& -p^2\,b_L\, g''/4
  \nonumber
\end{eqnarray}
Of course, in the mass eigenstate basis these terms must be supplemented with
the contributions from the gauge boson mixing. Recall that the mixing-induced
interactions are proportional to $1/g''$ and thus represent minor corrections.
We choose to evaluate the direct couplings at
the best-fit values of the free parameters $\Delta L$,
$\Delta R$, and $p$ when $g''=20$ and $\Lambda=1\unit{TeV}$ ($2\unit{TeV}$), 
i.e.\ $-0.0038$ ($-0.0039$), $0.0231$ ($0.0166$), and $0.14$ ($0.20$),
respectively.
Since the existing data restrict only the combinations of $b_{L,R}$
and $\lambda_{L,R}$ we will assume that $\lambda_{L,R}=0$.
The obtained coupling values are shown in Table~\ref{tab:cplngs}. If $\lambda$'s are allowed
to vary we expect that the best-fit couplings can double or reduce by half
without being accused of fine-tuning conspiracy between $b_{L,R}$'s and $\lambda_{L,R}$'s.}

\begin{table}[h]
\centering
\caption{\label{tab:cplngs}
         \blue{
         The couplings of the vector triplet to top/bottom quark vertices
         when $\lambda_{L,R}=0$ corresponding
         to the best-fit values of $\Delta L$, $\Delta R$, and
         $p$ when $g''=20$.
         }
        }
\begin{tabular}{ccccc}
\hline\noalign{\smallskip}
  $\Lambda$ & $g''b_L/4$ & $g''b_R/4$ & $p g''b_R/4$ & $p^2g''b_R/4$ \\
\noalign{\smallskip}\hline\noalign{\smallskip}
  1 TeV & -0.0190  & 0.1155 & 0.0162 & 0.0023 \\
  2 TeV & -0.0195  & 0.0830 & 0.0166 & 0.0033 \\
\noalign{\smallskip}\hline
\end{tabular}
\end{table}

In some models of partial fermion compositeness the masses of the
SM fermions are related to the product of compositeness
$\veps_{L,R}^f$ of the left and right
chirality~\cite{CHsketch,Sundrum,PomarolSerra}. The same
compositeness factors govern the strength of the couplings of the
new strong resonances to the fermions. Naively, we can relate
these considerations with the tBESS parameters as follows:
\begin{equation}
  \veps_L^{t,b} \propto b_L,\gap
  \veps_R^{t} \propto b_R,\gap
  \veps_R^{b} \propto p b_R.
\end{equation}
Then
\begin{equation}
  \frac{m_b}{m_t} = \frac{\veps_L^b\veps_R^b}{\veps_L^t\veps_R^t} = p.
\end{equation}
This example illustrates how the tBESS parameters can be related
to and fixed by predictions of specific models. In this particular
case, the predicted value of $p$ is about $0.03$. This is quite
away from the best fit $p\stackrel{>}{\sim} 0.2$ found in our
analysis. However, the best value of $p$ is not restricted very
tightly by the analyzed data; $p=0.03$ is still within the
$95\%$~CL region of the best fit by the four parameters.

If we fix $p=0.03$ as predicted
by some models with the partial fermion compositeness
then the best values of the remaining free parameters are $g''=24$,
$\Delta L=-0.004$, and $\Delta R=0.020$ when $\Lambda=1\unit{TeV}$.
Since $\chi^2 = 4.79$ the backing of this fit is about $9\%$.
If we assume $\Lambda=2\unit{TeV}$ then the numbers change as follows:
$g''=25$, $\Delta L=-0.004$, $\Delta R=0.014$, and $\chi^2 = 5.06$,
therefore the backing is about $8\%$.

%% file: 05-conclusions.tex

We have formulated and studied the effective Lagrangian 
for description of phenomenology of new scalar and vector resonances 
which might result from new strong physics beyond the SM.
Following the often used and studied approach the ESB sector of
the effective Lagrangian has been based on the 
$SU(2)_L\times SU(2)_R\rightarrow SU(2)_{L+R}$ non-linear 
sigma model while the scalar resonance has been introduced 
as the $SU(2)_{L+R}$ singlet and identified with the newly 
discovered 125-GeV boson. The vector resonance has been
built in as the $SU(2)_{L+R}$ triplet employing the hidden
local symmetry approach. Throughout the paper
we have assumed the vector resonance mass at the bottom
of the TeV scale.
No other non-SM fields have been 
considered in our effective Lagrangian.

Within this general framework we have invoked a special pattern
of interactions between the vector resonances and the SM fermions.
Beside the gauge boson mixing induced interactions
the symmetry of the Lagrangian admits the direct 
interactions of the vector triplet to the SM fermions.
Motivated by experimental as well as theoretical 
considerations we have opted for the pattern where
the vector resonance couples directly to the third quark generation
only. The couplings are chiral-dependent and 
the interaction of the right top quark can differ from that of 
the right bottom quark. Similar interaction patterns can be found 
in various recent extensions of the SM including extra-dimensional
and composite scenarios.

We have analyzed the tree-level unitarity of the gauge boson
scattering amplitudes to estimate the applicability
range of our effective Lagrangian. In particular,
we have been investigating how the presence of both, 
scalar and vector, resonances affects the unitarity 
restrictions. 
Adding the vector resonance triplet tends to improve
the unitarity when the coupling of the scalar to the gauge fields, $a$,
is lower than its SM value, $a=1$. For $a>1$ the presence of
the vector triplet further lowers the unitarity limit.
In general, the unitarity holds in the regions of $a$ and $g''$
that are also preferred by experiment. Recall that $g''$ is
the vector resonance triplet gauge coupling.

We have also analyzed the experimental limits on the free parameters
of the vector resonance triplet under the assumption of the SM
couplings for the scalar resonance. This was a simplifying assumption
that allowed us to focus on the vector resonance parameters before
any more complex and sophisticated investigation would be undertaken.
The assumption is in agreement with the current experimental findings
about the new 125-GeV boson. The results found in our analysis
could also be considered as an approximation of the situation when
the scalar resonance parameters slightly differ from their SM values.

\blue{
Since LHC measurements vertices are not restrictive
enough we have calculated the preferred values and CL intervals for
the vector resonance couplings from the low-energy observables only.
}
Namely, we have fitted five observables, $\epsilon_1$, $\epsilon_2$,
$\epsilon_3$, $\Gamma_b$, and $\mathrm{BR}(\bgs)$, parameterized
by four free parameters, $g''$, $\Delta L=b_L-2\lambda_L$, 
$\Delta R=b_R-2\lambda_R$, and $p$. 
When the cut-off scale $\Lambda=1\unit{TeV}$ the best-fit values read
$g''=29$, $\Delta L=-0.004$, $\Delta R=0.016$, and $p=0.209$ with
$\chi^2/\mathrm{d.o.f.}=2.40/1$ which corresponds to the $12\%$ backing.
The marginalized intervals of the $95\%$ CL region for individual
parameters read $g''\geq 12$, $-0.013 \leq \Delta L \leq 0.006$, 
$-0.006 \leq \Delta R \leq 0.056$. The marginalized interval for $p$
includes whole physically motivated interval $0\leq p\leq 1$.
Note that 
\blue{
while the best-fit value of $g''$ is somewhat above the naive
perturbativity limit
the perturbativity region overlaps with the found
restriction on $g''$.
}
Recall that the direct vector resonance coupling to the left
top-bottom quark doublet is proportional to $b_L g''$ and
the direct coupling to the right top quark is proportional
to $b_R g''$. With respect to the latter, the couplings of 
the right bottom quark to the charged and neutral vector 
resonances are diminished by $p$ and $p^2$, respectively. 
The parameters $\lambda_{L,R}$ parameterize the non-SM couplings
of the EW gauge bosons allowed by the symmetry.
With the low-energy measurements only one cannot obtain limits
on the $b$ and $\lambda$ parameters separately.

The best-fit value of $p$ found in our analysis seems to support
the assumption of some models of partial fermion compositeness
that the new strong physics resonances couple
stronger to the right top quark than to the right bottom quark. 
Unfortunately, when $g''$ is set
close to its most preferred value then any $p$ within
the $\langle 0;1\rangle$ interval has comparable experimental support.
Therefore, while the low-energy data seems to point in the right
direction any strong statements about the preferred value of $p$
would be premature at this point.
Unless the vector resonance is discovered directly,
further progress in the LHC measurements of the $Ztt$ and $Wtb$ vertices 
is needed to
improve limits on this and other parameters of the studied effective 
Lagrangian.

%% file: appendix-A.tex
\section{Experimental values}
\label{app:ExpValues}

In our analyses we have used the experimental values of the
epsilon pseudo-observables obtained from a fit to all LEP-I and
SLD measurements including the combined preliminary measurement of
the $W$-boson mass \cite{EpsilonData}:
\begin{eqnarray}
  \epsilon_1^{\mathrm{exp}} &=&(+5.4\phantom{4} \pm 1.0) \times 10^{-3},
  \label{epsExp1}\\
  \epsilon_2^{\mathrm{exp}} &=&(-8.9\phantom{4} \pm 1.2) \times 10^{-3},
  \label{epsExp2}\\
  \epsilon_3^{\mathrm{exp}} &=&        (+5.34 \pm 0.94) \times 10^{-3},
  \label{epsExp3}\\
  \epsilon_b^{\mathrm{exp}} &=&(-5.0\phantom{4} \pm 1.6) \times 10^{-3},
  \label{epsExpb}
\end{eqnarray}
with the correlation matrix
\begin{eqnarray}
\rho^\epsilon =
\left(%
\begin{array}{cccc}
  1.00 &  \phantom{-}0.60 & \phantom{-}0.86 &  \phantom{-}0.00 \\
  0.60 &  \phantom{-}1.00 & \phantom{-}0.40 & -0.01 \\
  0.86 &  \phantom{-}0.40 & \phantom{-}1.00 &  \phantom{-}0.02 \\
  0.00 & -0.01 & \phantom{-}0.02 &  \phantom{-}1.00 \\
\end{array}%
\right).
  \label{epsExpCM}
\end{eqnarray}

The value of the $Z\rightarrow b\bar{b}$ decay width
\begin{equation}
 \Gamma_b^{\mathrm{exp}} = (0.3773 \pm 0.0013)\ \mathrm{GeV}
\end{equation}
has been obtained from the experimental values~\cite{PDG2010}:
\begin{eqnarray}
 \mbox{BR}(Z\rightarrow b\bar{b})^{\mathrm{exp}} &=& (0.1512\pm 0.0005),
 \\
 \Gamma_{tot}(Z)^{\mathrm{exp}}\;\;\;\;\; &=& (2.4952 \pm 0.0023)~\mbox{GeV}.
\end{eqnarray}
The correlations between $\Gamma_b$ and $\epsilon_{1,2,3}$ have been
neglected.

For the branching fraction of $\bgs$ we have used the world
average \cite{btosgData} (CLEO, Belle, BaBar):
\begin{equation}
 \mbox{BR}(\bgs)^{\mathrm{exp}} = (3.55 \pm 0.26)\times 10^{-4}.
\end{equation}
We have considered no correlations between $\mbox{BR}(\bgs)$ and
any of the observables
$\epsilon_1,\epsilon_2,\epsilon_3,\Gamma_b$.

Below we will complete the list of numerical values that have been
used in the calculations of this paper:
\begin{eqnarray}
 \alpha(0)        &=& 1/137.036,
 \\
 \alpha(M_Z^2)    &=& 1/128.91,
 \\
 \alpha_s(M_Z^2)  &=& 0.1184,
 \\
 G_F  &=& 1.166364\times 10^{-5} \ \mathrm{GeV^{-2}},
 \\
 m_b  &=& 4.67 \ \mathrm{GeV},
 \\
 m_t  &=& 172.7 \ \mathrm{GeV},
 \\
 M_Z  &=& 91.1876 \ \mathrm{GeV},
 \\
 M_h  &=& 125 \ \mathrm{GeV}.
\end{eqnarray}
Then, using Eq.~(\ref{eq:GFinSM}) the numerical value of
$s_0^2$ is
\begin{equation}
 s_0^2 = 0.2311.
\end{equation}

%% file: appendix-B.tex
\section{Some anomalous couplings}
\label{app:AnomCplngs}

Here we show the anomalous couplings found
in the formulas~(\ref{eq:LariosEps1}) and (\ref{eq:BRb2gs})
for the loop contributions
$\epsilon_1^{\mathrm{vec}(1)}$ and $\mathrm{BR}(\bgs)$, respectively.
They read:
\begin{eqnarray}
 \kappa_L^{Wtb} &=& h(x;s_0)\left( 1-\frac{\Delta L}{2} \right)-1,
 \label{eq:kappaLWtbLE}\\
 \kappa_R^{Wtb} &=& h(x;s_0)\;\frac{p\;\Delta R}{2},
 \label{eq:kappaRWtbLE}\\
 \kappa_L^{Ztt} &=& -\frac{1}{2}\Delta L -\frac{4}{3}s_0^2\;\Delta k(x;s_0) ,
 \\
 \kappa_R^{Ztt} &=& +\frac{1}{2}\Delta R -\frac{4}{3}s_0^2\;\Delta k(x;s_0) ,
 \\
 \kappa_L^{Zbb} &=& +\frac{1}{2}\Delta L +\frac{2}{3}s_0^2\;\Delta k(x;s_0) ,
 \\
 \kappa_R^{Zbb} &=& -\frac{p^2}{2}\Delta R +\frac{2}{3}s_0^2\;\Delta k(x;s_0) ,
\end{eqnarray}
where $\Delta k(x;s_0)$ is given
in Eq.~(\ref{eq:DeltaRHOandKfromLEtBESS}) and
\begin{equation}
 h(x;s_0) = \frac{s_0}{s_\theta}\sqrt{\frac{1+4s_\theta^2 x^2}{1+x^2}}.
\end{equation}
The $x$ power expansion of this expression at $x=0$ reads
\begin{eqnarray}
 h(x;s_0) &=& 1-s_0^2\;\Delta k(x;s_0) + \cO(x^4)
 \nonumber\\
 &=& 1-0.430 \,x^2 - 0.405 \,x^4 +\ldots .
\end{eqnarray}
